\documentclass[10pt]{article}      

\usepackage[reqno,centertags]{amsmath}
\usepackage{amsfonts,amssymb,ifthen,cite}

 \makeatletter

  \renewcommand\section{\bigskip\@startsection {section}{1}{\z@}%
                       {-3.5ex \@plus -1ex \@minus -.2ex}%
                       {2.3ex \@plus.2ex}%
                       {\reset@font\centering\Large\bfseries}}
  \renewcommand\subsection{\@startsection{subsection}{2}{\z@}%
                       {-3.25ex\@plus -1ex \@minus -.2ex}%
                       {1.5ex \@plus .2ex}%
                       {\reset@font\centering\large\bfseries}}

   \ifthenelse{\@ptsize = 0}%
    {\newcommand{\openone}{\leavevmode\hbox{\small1\kern-3.3pt\normalsize1}}}%
    {\newcommand{\openone}{\leavevmode\hbox{\small1\kern-3.8pt\normalsize1}}}

   \newcommand{\opone}{{\hat{\openone}}}

   \DeclareSymbolFont{boperators}{OT1}{cmr}{bx}{n}
   \DeclareMathSymbol{\nordl}{\mathopen}{boperators}{"3A}
   \DeclareMathSymbol{\nordr}{\mathclose}{boperators}{"3A}

   \DeclareMathOperator{\sign}{sign}
   \DeclareMathOperator{\detix}{det}
   \DeclareMathOperator{\Det}{Det}
   \DeclareMathOperator{\tr}{tr}
   \DeclareMathOperator{\Tr}{Tr}
   \DeclareMathOperator{\re}{Re}
   \DeclareMathOperator{\im}{Im}

  \newcounter{note}

  \newcommand{\makenotemark}[1]{\mbox{$^{#1}$}}
  \newcommand{\labelnote}{\makenotemark\thenote}

  \newcommand{\noteitem}[1]{\item\label{#1}}
  \newcommand{\notesname}{Notes}

  \newenvironment{notes}
    { \section*{\notesname}%
      \begin{list}{\labelnote}{\usecounter{note}}%
      }
    { \end{list}%
      }

  \newcommand{\note}[1]{\makenotemark{\ref{#1}}}

  \numberwithin{equation}{section}

  \newcommand{\abs}[1]{\lvert#1\rvert}

  \newcommand{\grad}{{\mathrm{d}}}
  
  \newcommand{\normalvect}{{\vec{n}}}
  \newcommand{\domain}{{\mathit\Omega}}
  
  \newcommand{\Chi}{{\mathcal X}}

  \newcommand{\variatb}{{\boldsymbol{\delta}}}
  \newcommand{\qPhi}{{\hat{\Phi}}}

  \newcommand{\absidx}{\boldsymbol}

  \newcommand{\fctsct}{{\mathfrak{F}}}

  \newcommand{\dnstsct}{{\tilde{\mathfrak{F}}}}

  \newcommand{\gdnst}{{{\mathfrak g}^{\frac12}}}
  
  \newcommand{\qdnst}{{{\mathfrak q}^{\frac12}}}

  \newcommand{\hist}{{\boldsymbol{\mathsf H}}}
  \newcommand{\phasesp}{{\boldsymbol{\Gamma}}}
  \newcommand{\quantsp}{{\mathcal{H}}}
  \newcommand{\quantobs}{{\mathsf{L}(\quantsp)}}

  \newcommand{\distdl}{{\stackrel{\leftharpoonup}{\mathrm{d}}}}
  \newcommand{\distdr}{{\stackrel{\rightharpoonup}{\mathrm{d}}}}

  \newcommand{\distdFr}{{\stackrel{\rightharpoonup}{d F}}}
  \newcommand{\distdFl}{{\stackrel{\leftharpoonup}{d F}}}
  \newcommand{\symplstr}{{\partial F}}
  
  \newcommand{\Boxr}{{\stackrel{\rightharpoonup}{\square}}}

  \newcommand{\spcin}{\quad}
  \newcommand{\spcpnct}{\quad}
  \newcommand{\spce}{\spcpnct\:}
  \newcommand{\period}{{\mbox{\spcpnct.}\relax}}
  \newcommand{\comma}{{\mbox{\spcpnct,\spcpnct}\relax}}
  \newcommand{\commae}{{\mbox{\spcpnct,}\relax}}

  \newcommand{\AoperH}{{\mathsf A}}
  \newcommand{\BmetrH}{{\mathsf B}}
  \newcommand{\CmetrH}{{\mathsf C}}
  \newcommand{\DoperH}{{\mathsf D}}

  \newcommand{\cauchyproj}{\boldsymbol{\iota}}

  \newcommand{\traI}{{\mathrm I}}
  \newcommand{\traV}{{\mathrm V}}
  \newcommand{\traL}{{\mathrm \Lambda}}

  \newcommand{\GFC}{{G_C}}
  \newcommand{\GFA}{{G_{adv}}}
  \newcommand{\GFR}{{G_{ret}}}
  \newcommand{\GFRA}{{G_{ret,adv}}}
  \newcommand{\GFS}{{\bar{G}}}
  \newcommand{\GFH}{G^{(1)}}
  \newcommand{\GFF}{G^F}
  \newcommand{\GFP}{G^+}
  \newcommand{\GFM}{G^-}
  \newcommand{\GFPM}{G^\pm}
  \newcommand{\GFMP}{G^\mp}

  \newcommand\frqp[2]{\mbox{\kern-2pt$\genfrac{}{}{0pt}{1}{#1}{#2}$}}

 \makeatother

\begin{document}

 \title{\bfseries Particle Interpretations and Green Functions for a Free
 Scalar Field\thanks
        {There exists a broader version \protect\cite{Krtous:pigflong} of
        this paper with additional review material.}
        }

 \author{\textbf{Pavel Krtou\v{s}}\thanks
         {\protect\parbox[t]{\linewidth}{
               e-mail: \texttt{krtous@phys.ualberta.ca},\protect\\
               www:
               \texttt{http://fermi.phys.ualberta.ca/\~{}krtous/HomePages/}
               }
         }
         \medskip\\
         \small{412 Avadh Bhatia Phys. Lab.,}\\
         \small{University of Alberta,}\\
         \small{Edmonton, Alberta T6G2J1,}\\
         \small{Canada}%
         }

 \date{July 11, 1995}

 \maketitle

 \begin{abstract}
 The formalism of Ashtekar and Magnon \cite{AshtekarMagnon:1975} for the
 definition of particles in quantum field theory in curved spacetime is
 further developed. The relation between basic objects of this formalism
 (e.g., the complex structure) and different Green functions is found. It
 allows one to derive composition laws for Green functions.

 The relation of two definitions of particles is reformulated in the
 formalism and the base-independent Bogoljubov transformation is expressed
 using quantities which are derivable directly from the ``in-out'' Green
 function.

 \end{abstract}

 \newpage


\section*{Introduction}

In this paper we will investigate the notion of particles in quantum field
theory in curved background and the properties of different Green functions.
Quantum field theory in curved spacetime is a well known theory developed
already two decades ago \cite{DeWitt:1975} (see also
\cite{BirrellDavies:book,Fulling:book,Gribetal:book,Wald:book1994}). We will
not discuss any new physical effects, but we will develop a useful
mathematical framework based on \cite{AshtekarMagnon:1975} which, for
example, allows us to derive a lot of nontrivial relations among Green
functions and helps us to understand the connection of different notions of
particles. Beside making a reformulation of many known facts (such as the
possibility of the reconstruction of initial and final particle states from
the in-out Green function \cite{RumpfUrbantke:1978}) in this framework, we
will show a relation between the complex structure used for the definition of
particles and the Hadamard Green function, and we will express a
base-independent Bogoljubov transformation in terms which are derivable
directly from the in-out Green function --- facts which we believe are not
published elsewhere.

This paper is composed of three parts. In part \ref{sec:GD} the structure of
covariant phase-space is introduced. It is based mainly on the formulation by
DeWitt \cite{DeWitt:book1965}.

Part \ref{sec:partint} defines particle representations of quantum algebra.
The standard approach for the definition of particles is a decomposition of
the quantum field into modes --- what essentially expresses the field as a
system of harmonic oscillators. This approach could be called a ``brute
force'' approach --- it depends on a choice of a base, partially in
unimportant, but partially in very important, ways. There exists another,
more formal, way to define particles, introduced by Ashtekar and Magnon
\cite{AshtekarMagnon:1975}. They have used a ``complex structure'' on the
phase space of a classical system to define particles. We will further
develop this formulation, and we will show a close relation between the
complex structures (and their projector operators) and different kinds of
Green functions. Next we will show that a condition of diagonalization of a
Hamiltonian in a particle base picks up uniquely a notion of particles. This
is a slightly different condition than the one used in
\cite{AshtekarMagnon:1975} but leads to the same particle interpretation.
Finally the boundary conditions for the Feynman Green function are found (see
also \cite{RumpfUrbantke:1978}), and different composition laws for Green
functions are derived.

The last part investigates a relation between two particle interpretations
which is important for scattering situations. The developed formalism allows
one to define a base independent Bogoljubov transformation and using it
easily find a canonical base in which Bogoljubov coefficients are diagonal
\cite{Hajicek:1977}. Operators $\alpha_o$, $\beta_o$ on the phase space which
plays a role of Bogoljubov coefficients can be expressed using a single
operator $\Chi$ as $\alpha_o = \cosh\Chi$, $\beta_o = \sinh\Chi$ where the
operator $\Chi$ appears in many important quantities as an in-out Green
function, S-matrix or transition amplitudes. In the part \ref{sec:inoutform}
we also show that initial and final notions of particles are possible to
reconstruct from a knowledge of a Green function which satisfies certain
conditions (see also \cite{RumpfUrbantke:1978}). So, if we are able to
construct such a Green function using other methods as, for example, a path
integral, this result allows us to define a corresponding notion of particles
in standard quantum theory. Finally we will find boundary conditions for the
in-out Feynman Green function and other composition laws among Green
functions.

\section{General definitions}
\label{sec:GD}


\subsection*{Covariant phase space}

We will investigate a free scalar field theory. The \emph{space of histories}
$\hist$ (i.e., field configurations in spacetime) for a real scalar field is
the space of real functions\note{bundlesnotation} on a spacetime manifold
$M$. It is a vector space, and a vector index of $\phi^x$ denotes essentially
a point $x$ in spacetime\note{positasindex}, which we will usually suppress.
A dot ``$\bullet$'' will denote a summing over these ``$\infty^4$'' indexes,
i.e. integration over the spacetime $M$.

The wave operator on $\hist$ is
\begin{equation}
  F = - \distdl_{\absidx{\alpha}}
  \bullet (g^{\absidx{\alpha\beta}} \gdnst \delta) \bullet
  \distdr_{\absidx{\beta}} -
  (V \gdnst \delta) = (\gdnst \Boxr) - (V \gdnst \delta)
  \spcin\in\;\hist_2^0\period \label{sfF}
\end{equation}
Here $g_{\absidx{\alpha\beta}}$ is the spacetime metric\note{indexes},
$\gdnst = (\Det g)^{\frac12} \in \dnstsct\,M$ is the metric volume element
and $V$ is a spacetime dependent potential\note{distributions}.

The equation of motion is
\begin{equation}\label{freeeqofmot}
  F \bullet \phi = 0\period
\end{equation}
We will call the space of solutions a \emph{covariant phase space}
$\phasesp$. The phase space $\phasesp$ is generally nonlinear, but for free
field it forms a vector space. The vector index labels Cauchy data for the
equation of motion (for example value and momentum of field on a Cauchy
hypersurface). The dot ``$\circ$'' will represent contraction over these
``$2\infty^3$'' indexes.

We will introduce a symplectic structure on $\phasesp$ using the Wronskian
$\symplstr[\Sigma]$ of the operator $F$. We have for any spacetime domain
$\domain = \langle\Sigma_f,\Sigma_i\rangle$ between two Cauchy hypersurfaces
$\Sigma_i$, $\Sigma_f$
\begin{align}
\begin{split}\label{Fchicomm}
  \symplstr[\partial\domain] &=
  (\chi[\domain]\,\delta) \bullet F - F \bullet (\chi[\domain]\,\delta)
  \spcin\in \hist_2^0\commae \\
  \symplstr[\partial\domain] &= \symplstr[\Sigma_f] - \symplstr[\Sigma_i]
  \commae
\end{split}\\
  \symplstr[\Sigma] &= \distdFl[\Sigma] - \distdFr[\Sigma]  \commae
  \label{defofsymplstr}
\end{align}
where $\chi[\domain]$ is a characteristic function of the domain $\domain$
and
\begin{equation}\label{sfdFr}
  \distdFr[\Sigma] =
  (\gdnst\,\delta[\Sigma]\,\normalvect^{\absidx{\alpha}})
  \distdr_{\absidx{\alpha}} \comma
  \distdFl[\Sigma] = \distdFr[\Sigma]^{\top} =
  \distdl_{\absidx{\alpha}}
  (\gdnst\,\delta[\Sigma]\,\normalvect^{\absidx{\alpha}}) \period
\end{equation}
Here $\normalvect^{\absidx{\alpha}}$ is a future oriented normal to the
hypersurface $\Sigma$, and $\delta[\Sigma]$ defined by
\begin{equation}
  \delta[\partial\domain] = \delta[\Sigma_f] - \delta[\Sigma_i] =
  - \normalvect^{\absidx{\alpha}} (\grad_{\absidx{\alpha}} \chi[\domain])
\end{equation}
is a delta function localized on the hypersurface $\Sigma$, i.e. for
$\psi\in\fctsct\,M$
\begin{equation}
  \int_{M} { \psi\,\delta[\Sigma]\,\gdnst} = \int_{\Sigma} {\psi\,\qdnst}
  \commae
\end{equation}
where $\qdnst\in\dnstsct\,\Sigma$ is the volume element on the hypersurface
$\Sigma$.

The Wronskian defines the usual Klein-Gordon product
\begin{equation}
  \phi_1  \bullet  \symplstr[\Sigma] \bullet \phi_2
   = \int_{\Sigma} \bigl( (\normalvect^{\absidx{\alpha}}
   \grad_{\absidx{\alpha}} \phi_1) \phi_2 -
  \phi_1 (\normalvect^{\absidx{\alpha}} \grad_{\absidx{\alpha}} \phi_2)
  \bigr) \qdnst \period
\end{equation}
We have for $\phi_1, \phi_2\in\phasesp$, using (\ref{Fchicomm}) and
(\ref{freeeqofmot})
\begin{equation}
  \phi_1 \bullet \symplstr[\Sigma_f] \bullet \phi_2
  - \phi_1 \bullet \symplstr[\Sigma_i] \bullet \phi_2 =
   \phi_1 \bullet \bigl( (\chi[\domain]\,\delta) \bullet F
  - F \bullet (\chi[\domain]\,\delta) \bigr) \bullet \phi_2 = 0 \period
\end{equation}
So $\symplstr[\Sigma]$ is independent of $\Sigma$ if it acts on vectors from
$\phasesp$. Therefore we can introduce a restriction $\symplstr \in
\phasesp_2^0$ of $\symplstr[\Sigma] \in \hist_2^0$ which is not dependent on
a hypersurface $\Sigma$. $\symplstr$ is an antisymmetric nondegenerated
closed bi-form on $\phasesp$, which means it is a symplectic structure of our
phase space.

We can define its inversion\note{operatorsconv}
\begin{equation}
  \GFC \in \phasesp_0^2  \comma  \GFC^{\top}  =  - \GFC \comma
  \GFC \circ \symplstr = \symplstr  \circ  \GFC = -
  \delta^{(\phasesp)}\spce\label{defGFCinphasesp}
\end{equation}
or, formulated on the space of histories $\hist$,
\begin{gather}
  \GFC \in \hist_0^2
  \comma  F \bullet \GFC = \GFC \bullet F = 0 \comma
  \GFC^{\top} =  - \GFC\commae\nonumber\\
  \GFC \bullet \symplstr[\Sigma] = \symplstr[\Sigma]  \bullet  \GFC = -
  D_C[\Sigma]\period
\end{gather}
Here $\delta^{(\phasesp)}$ is the identity operator on $\phasesp$ and
$D_C[\Sigma]$ propagates Cauchy data on a hypersurface $\Sigma$ to a solution
of the equation of motion (\ref{freeeqofmot}). It is a projection from
$\hist$ to $\phasesp$ using the Cauchy data on $\Sigma$. It induces a mapping
between spaces $\hist_k^l$ and $\phasesp_k^l$ which we will call
$\cauchyproj[\Sigma]$.

$\GFC$ is called the causal Green function. It is possible to show (see
\cite{DeWitt:book1965}) that
\begin{equation}
  \GFC = \GFR - \GFA  \commae
\end{equation}
where $G_{ret}, G_{adv}$ are retarded and advanced Green functions for
equation (\ref{freeeqofmot}).

The Poisson bracket of two observables $A, B \in \fctsct\,\phasesp$ is given
by\note{gammavariat}
\begin{equation}
  \label{poissonbr}
  \{ A, B \} = \variatb\,A \circ \GFC \circ \variatb\,B \period
\end{equation}
Applied to the basic variable $\Phi^x$ --- a value of the field at a point
$x$, we get
\begin{equation}\label{sfcommrel}
  \{\Phi^x,\Phi^y\} = \GFC^{xy}\period
\end{equation}

The definitions of the symplectic structure and the causal Green function
give us
\begin{equation}
  \phi = -\,\GFC \bullet \symplstr[\Sigma] \bullet \phi =
  -\,\GFC \bullet \distdFl[\Sigma] \bullet \phi + \GFC \bullet
  \distdFr[\Sigma] \bullet \phi
\end{equation}

Using the fact that $\distdFl[\Sigma]_{xy}$ as a function of $y$ has a
support on $\Sigma_t$ and does not include any derivatives in directions out
of $\Sigma$ and that
\begin{equation}
  \GFC^{xy} = 0 \quad\text{for}\quad x,y \in \Sigma\commae
\end{equation}
we see that the operator
\begin{equation}\label{Dphidef}
  D_\varphi[\Sigma] = -\,\GFC \bullet \distdFl[\Sigma]
\end{equation}
plays a role of a projector on the subspace of solutions with zero normal
derivative on $\Sigma$, and the operator
\begin{equation}\label{Dpidef}
  D_\pi[\Sigma] = \GFC \bullet \distdFr[\Sigma]
\end{equation}
is a projector on a subspace of solutions with zero value on $\Sigma$. Let's
summarize the properties of the operators $D_\varphi[\Sigma]$ and
$D_\pi[\Sigma]$:
\begin{gather}
  D_\varphi[\Sigma]\bullet D_\varphi[\Sigma]
  = D_\varphi[\Sigma]\comma D_\pi[\Sigma]\bullet
  D_\pi[\Sigma]=D_\pi[\Sigma]\commae\nonumber\\
  D_\varphi[\Sigma]\bullet D_\pi[\Sigma] =  D_\pi[\Sigma]\bullet
  D_\varphi[\Sigma]=0\commae\label{Dphipiproj}\\
  D_\varphi[\Sigma] + D_\pi[\Sigma] = D_C[\Sigma]\period\nonumber
\end{gather}

We can map tensors $\distdFl[\Sigma], \distdFr[\Sigma], D_\varphi[\Sigma],
D_\pi[\Sigma]$ to spaces $\phasesp^k_l$ using the map $\cauchyproj[\Sigma]$.
We will use the same symbols for the result. The definitions (\ref{Dphidef}),
(\ref{Dpidef}) of $D_\varphi[\Sigma]$, $D_\pi[\Sigma]$ and relations
(\ref{Dphipiproj}) can be rewritten as
\begin{gather}
  \symplstr = D_\varphi[\Sigma]\circ\symplstr\circ D_\pi[\Sigma]
  + D_\pi[\Sigma]\circ\symplstr\circ D_\varphi[\Sigma]\commae\\
  \GFC =  D_\varphi[\Sigma]\circ\GFC\circ D_\pi[\Sigma]
  + D_\pi[\Sigma]\circ\GFC\circ D_\varphi[\Sigma]\commae\\
  \distdFl[\Sigma] = D_\pi[\Sigma]\circ\distdFl[\Sigma]\circ
  D_\varphi[\Sigma]\comma
  \distdFr[\Sigma] = D_\varphi[\Sigma]\circ\distdFr[\Sigma]\circ
  D_\pi[\Sigma]\period
\end{gather}
We will also introduce the ``inversions'' of bi-forms $\distdFr[\Sigma],
\distdFl[\Sigma]\in\hist^0_2$
\begin{equation}
\begin{split}
  \distdFr[\Sigma]^{-1}\bullet\distdFr[\Sigma] = D_\pi[\Sigma]&\comma
  \distdFr[\Sigma]\bullet\distdFr[\Sigma]^{-1} = D_\varphi[\Sigma]\commae\\
  \distdFl[\Sigma]\bullet\distdFl[\Sigma]^{-1} = D_\pi[\Sigma]&\comma
  \distdFl[\Sigma]^{-1}\bullet\distdFl[\Sigma] = D_\varphi[\Sigma]\spce
\end{split}
\end{equation}
and similarly for $\distdFr[\Sigma], \distdFl[\Sigma]\in\phasesp^0_2$.

Using the covariant phase space $\phasesp$ and a solution $\Phi$ of Eq.\
(\ref{freeeqofmot}) as a basic variable is nothing other than the
``Heisenberg picture'' in classical physics.


\subsection*{Quantization}

Quantization is a heuristic procedure of construction of a quantum theory for
a given classical theory. Let us have a classical system described by a phase
space $\phasesp$ and symplectic structure $\symplstr$. Observables are
functions on $\phasesp$, and the Poisson bracket is given by
(\ref{poissonbr}). Quantization tells us to assign to at least some
observables quantum observables --- operators on a quantum Hilbert space
$\quantsp$. Let denote the space of operators $\quantobs$. We will use
letters with a hat to denote operators. The quantum observables should
satisfy the same algebraic relations as the classical ones and commutation
relations generated by Poisson brackets. If the quantum versions of classical
observables $A,B$ and $C = \{A,B\}$ are $\hat{\mathrm A},\hat{\mathrm
B},\hat{\mathrm C}$, they should satisfy
\begin{equation}
  [\hat{\mathrm A},\hat{\mathrm B}] = \hat{\mathrm A}\hat{\mathrm
  B}-\hat{\mathrm B}\hat{\mathrm A} = -i \hat{\mathrm C}\period
\end{equation}

It is well known that the procedure described above is not possible to carry
out for all classical observables. Because quantum observables do not commute
we have  an ``ordering problem'' for observables given by a product of
noncommuting observables.

For a scalar field we would like to define a quantum theory as the using
basic observable $\Phi^x$ --- a ``value of field at point $x$''. Using Eq.\
(\ref{sfcommrel}) we can write the commutation relations for $\qPhi^x$ as
\begin{equation}
  [\qPhi^x , \qPhi^y ] = - i \hat{G}_C^{xy}  \commae
\end{equation}
where $\hat{G}_C^{xy}$ is a quantum version of the causal Green function
$\GFC$. But for an interacting field $\GFC$ does depend on $\Phi$ (position
in $\phasesp$). This means that to define $\hat{G}_C^{xy} = \GFC(\qPhi)$ we
have to solve the ordering problem. This is generally a nontrivial task.

This problem can be to solved for linear theories, for example, that of a
free scalar field. In this case the causal Green function $\GFC$ is constant
on $\phasesp$ (independent of $\Phi$). Therefore its quantum version is
proportional to the unit operator $\opone$ (Poisson brackets of any
observable with a constant are zero, so the quantum version of a constant
observable has to commute with all observables). So
\begin{equation}\label{qPhicomrel}
  [\qPhi^x, \qPhi^y] = - i \GFC^{xy} \opone\period
\end{equation}
This means that we do not have any ordering problem in the quantization of
observables $\Phi^x$. These observables are linear on $\phasesp$, and any
linear observable can be generated from them without the multiplication of
two noncommuting observables. So we have a unique quantization of all linear
observables on the phase space $\phasesp$.

For quadratic observables
\begin{equation}
  A(\Phi) = \frac12 \Phi\circ a \circ\Phi\comma a\in\phasesp_2^0 \commae
\end{equation}
we have an ambiguity in the ordering of two $\qPhi$ observables. However,
because the commutator of two $\qPhi$ is proportional to the unit operator,
any quantum version of $A(\Phi)$ can be written as
\begin{equation}
  \hat{\mathrm A} = \frac12 \qPhi \circ a \circ \qPhi + \alpha \opone
  \commae
\end{equation}
where the factor $\alpha$ is given by a particular choice of an operator
ordering.


\subsection*{Infinite dimension of the phase space}

We are working with $\phasesp$ and $\quantsp$ on very formal level. More
precisely using structures on $\phasesp$ and a quantization scheme, we define
a quantum algebra of observables (generated by an algebra valued distribution
$\qPhi$) which we want to represent using operators on $\quantsp$. The
representation is unique (up to unitary transformations) in the case of a
finite dimensional phase space $\phasesp$. But there exist unitarily
inequivalent representations of the quantum algebra on $\quantsp$ in the case
of a infinite dimensional phase space $\phasesp$. The phase space of the
scalar field theory is infinite dimensional, and we really will deal with
inequivalent representations.

We will adopt the following intuitive point of view. We will look on
$\quantsp$ as a vector space with a ``quantum product'' between any two
states (vectors). However, there exist sets of states with a nondegenerated
quantum product for any two states from the same set but maybe a degenerated
quantum product for states from different sets.  Each of these sets with
restriction of the quantum product on it forms a Hilbert space, but there is
no well defined unitary operator in this Hilbert space which relates one set
to other set.

However it is useful keep all these sets together in one quantum space
$\quantsp$ with ``generalized'' quantum product. Intuitively these sets
represents different ``phases'' of the same physical system. Physically a
vacuum of one phase will contain an infinite mean number of particles of
other phase.

For a complete discussion of inequivalent representation see
\cite{Wald:1979a,Wald:book1994}.


\section{Particle interpretation}
\label{sec:partint}


\subsection*{What are particles and why we need them?}

In this part we would like to discuss  quantization of the free scalar field
in more detail. We have seen that the algebra of quantum observables is
generated by basic observable $\qPhi$ with commutation relations
(\ref{sfcommrel}), and we have a unique quantum version of any classical
linear observable. For quantization of nonlinear observables we have to
choose a particular operator ordering.

But we would like to find more about the structure of the space $\quantsp$.
For the interpretation of a theory we usually need to pick up quantities
which are measurable in a physical experiment. This means one needs to find
quantities to which a realistic detector coupled to the field is sensitive.
But we need even more for understanding a theory. We would like to have an
intuitive picture, a more descriptive way how to speak about our quantum
system.

A very useful way to describe a quantum field is a language of particles. It
is possible to construct special states representing ``definite numbers of
quanta of field'', and the structure of the theory becomes often much simpler
when it is expressed in terms of these states. Particle states do not have to
be always straightforwardly measurable quantities --- only in special
situations it is possible to construct a simple detector sensitive exactly to
some particle states. But even in situations when it is not simple to prepare
the system in a particle state, it can be useful to use such states for
description of physical processes.

First we have to define what particles are. Here is a short list of some
elementary properties of particles:
\begin{list}{\textbf{-}}{%
\itemsep=0pt%
\parsep=0pt}
\item discrete nature of particles
\item particles as quanta of energy
\item definiteness of position or momenta of particles
\item measurability by a detector
\item a connection with quantization of a relativistic particle
\end{list}

The first property is the property we will use the most. Particles can be
counted; they have a piece-like character. We speak about photons because we
are able to detect discrete hits on a screen when we illuminate it by a weak
electromagnetic field. We speak about quanta of energy in the case of a
hydrogen atom because the atom can emit the energy in discrete pieces.

A discrete nature is only one of many properties of classical particles. But
we are not speaking about classical particles. We want to construct a useful
notion of particles for quantum field theory. And the notion of discrete
pieces gives us the weakest sense of the word particle. Or, maybe, it would
be better to speak about quanta of the field.

Of course, we can be more restrictive about the notion of particles. As the
second item in our list suggests, we could require that particles are quanta
of energy - i.e. that some particle states are eigenstates of a Hamiltonian
of our system. We will discuss this condition later. Let us only say that
this condition picks up a unique notion of particles but, unfortunately, in a
general situation we do not have a unique notion of energy.

Similarly, it is difficult to give a well defined sense to position or
momentum of a particle in a general spacetime. Only in the case when our
spacetime is sufficiently special we can introduce some generalized momentum
operator or to find a simple detector sensitive exactly to some kind of
particles \cite{Unruh:1976,BirrellDavies:book}. Localizability is an even
more subtle issue.

There exists a complete different way how to construct quantum field theory
\cite{DeWitt:1975,Hartle:LH1992,Krtous:inprep}. It is possible to quantize a
relativistic particle using a sum over histories approach and to find that
transition amplitudes calculated in this way are exactly the same as
amplitudes between some particle states of the scalar field theory
\cite{Krtous:inprep}. It gives us another interesting meaning to particle
states.


\subsection*{Fock structure of the quantum space $\quantsp$.}

Now let us concentrate on the basic particle property --- the discrete nature
of particles. We would like to speak about a state with no particles, about
states with one particle, two particles and so on. The representation of this
structure in quantum mechanics is well known. We want to find a Fock
structure in our quantum space $\quantsp$ which divides $\quantsp$ to
subspaces with a vacuum state, one, two and more particle states. As known,
the Fock structure can be generated by creation and annihilation operators
$\hat{\mathrm a}_{\mathrm{k}}^\dagger$ and $\hat{\mathrm a}_{\mathrm{k}}$
which satisfy the commutation relations\note{onlybosons}
\begin{equation}
  [ \hat{\mathrm a}_{\mathrm{k}} , \hat{\mathrm a}_{\mathrm{l}} ] = 0\comma
  [ \hat{\mathrm a}_{\mathrm{k}}^\dagger , \hat{\mathrm
  a}_{\mathrm{k}}^\dagger ] = 0\comma
  [ \hat{\mathrm a}_{\mathrm{k}} , \hat{\mathrm a}_{\mathrm{l}}^\dagger ] =
  \alpha_{\mathrm{kl}}\,\opone  \commae
\end{equation}
where indexes $\mathrm{k,l}$ label one-particle states and
$\alpha_{\mathrm{kl}}$ is a transition amplitude between two one-particle
states labeled by $\mathrm{k}$ and $\mathrm{l}$. So, to find a particle
interpretation of the free scalar field theory we need to construct such
creation  and annihilation operators from our basic observable $\qPhi$. The
construction which we will describe below is possible to find in
\cite{AshtekarMagnon:1975}.

First we describe a way in which we will label one-particle states. Particles
used for the description of a scalar field are scalar particles without any
inner degrees of freedom. This means that on the classical level the position
and momenta at one time are sufficient for the determination of the state of
one particle.  Therefore the quantum space $\quantsp_1$ of one-particle
states should be ``$\infty^3$'' dimensional (one vector of a base for each
space point) as a \emph{complex} vector space (let's denote it $\mathbb
C$-dimensionality). It means that as real vector space it has ``the same''
$\mathbb R$-dimension as the phase space $\phasesp$ of the scalar field ---
``$2\infty^3$''. These formal consideration suggest the use of the space
$\phasesp$ for labeling of one-particle states. More precisely, we would like
to find a one-to-one map between spaces $\quantsp_1$ and $\phasesp$.

This faces us with a problem. Some vectors in $\quantsp_1$ are related only
by a phase and essentially represent the same physical state. But their
images (labels) in $\phasesp$ are different. We would like to know how these
different labels are related. This means that we need to introduce a
structure of a Hilbert space to the phase space $\phasesp$ in such way that
our mapping between $\phasesp$ and $\quantsp_1$ will be an isomorphism of
Hilbert spaces.

For this we need to speak about $\phasesp$ as a complex vector space with a
positive definite scalar product. Therefore we need to define ``a
multiplication by a complex number'' in $\phasesp$. We do not want to
complexify $\phasesp$ to $\phasesp^{\mathbb C} = {\mathbb C}\otimes \phasesp$
because it would ``doubled'' the dimension. We need a multiplication by a
complex number \emph{inside} of $\phasesp$.

To do it, it is sufficient to define multiplication by an imaginary unity.
Let define for $\phi\in\phasesp$
\begin{equation}
  i \star \phi = J \circ \phi  \commae
\end{equation}
where ``$\star$'' represents multiplication of a vector from $\phasesp$ by a
complex number which we are defining and $J$ is an operator on $\phasesp$. It
is clear that for  consistency $J$ has to be a linear operator (it is already
reflected in using ``$\circ$'' operation) which satisfies\note{phasespdelta}
\begin{equation}\label{Jsquare}
  J \circ J = - \delta\period
\end{equation}
Such operator is called a complex structure on the vector space $\phasesp$.
There exist a lot of different complex structures on $\phasesp$, and we will
discuss their relation in part \ref{sec:inoutform}. At this moment we will
pick up one which we denote $J_p$, and we will use the index $p$ for all
quantities which depend on this complex structure.

Next we need to define a positive definite product on $\phasesp$. It is
possible to do if we will assume that $J_p$ possess the following properties
\begin{gather}
  J_p \circ \symplstr \circ J_p = \symplstr \quad \Leftrightarrow \quad
  J_p \circ \symplstr = - \symplstr \circ J_p \commae
  \label{JdFcompatibility}\\
  \gamma_p = J_p \circ \symplstr \quad\text{is positive definite}\period
  \label{JdFpositivity}
\end{gather}
The first property is called compatibility of $J_p$ with the symplectic
structure $\symplstr$. Let's note that if $J_p$ is compatible with
$\symplstr$, the bi-form $\gamma_p$ is automatically symmetric. Let assume
that $J_p$ satisfies both conditions (\ref{JdFcompatibility}) and
(\ref{JdFpositivity}). Now we can define a scalar product on $\phasesp$ by
\begin{equation}\label{scalarproduct}
  \langle \phi_1 , \phi_2 \rangle_p = \phi_1 \circ \frac12 (\gamma_p - i
  \symplstr) \circ \phi_2\period
\end{equation}
It is linear in the second argument, antilinear in the first one and positive
definite.
\begin{gather}
  \langle i \star \phi_1 , \phi_2 \rangle_p = -i
  \langle\phi_1,\phi_2\rangle_p\comma
  \langle\phi_1,i\star\phi_2\rangle_p = i
  \langle\phi_1,\phi_2\rangle_p\commae\\
  \langle\phi,\phi\rangle_p = \frac12 \phi\circ\gamma_p\circ\phi > 0
  \quad\text{for}\quad \phi\not=0\period
\end{gather}

We finally changed the phase space $\phasesp$ to a Hilbert space $\phasesp_p
= (\phasesp,\star,\langle\,,\,\rangle_p)$ using a new extra structure $J_p$.
Now we can proceed and define creation and annihilation operators acting on
$\quantsp$ which are labeled by vectors from $\phasesp$:
\begin{equation}
  \hat{\mathrm a}_p[\phi] = \langle \phi, \qPhi \rangle_p\comma
  {\hat{\mathrm a}_p[\phi]}^\dagger = \langle \qPhi, \phi \rangle_p\spce
\end{equation}
for any $\phi\in\phasesp$. We will show below (see Eq.\
(\ref{anihcreatommrel})) that the commutation relations of such defined
operators are
\begin{gather}
  \bigl[\hat{\mathrm a}_p[\phi_1],\hat{\mathrm a}_p[\phi_2]\bigr] = 0 \comma
  \bigl[{\hat{\mathrm a}_p[\phi_1]}^\dagger , {\hat{\mathrm
  a}_p[\phi_2]}^\dagger\bigr] = 0\commae\nonumber\\
  \bigl[\hat{\mathrm a}_p[\phi_1],{\hat{\mathrm a}_p[\phi_2]}^\dagger\bigr] =
  \langle \phi_1,\phi_2\rangle_p\,\opone
  \period \label{ancrcomrel}
\end{gather}
We see that they really satisfy the commutation relations of creation and
annihilation operators. We can define a vacuum state by the condition
\begin{gather}
  \hat{\mathrm a}_p[\phi] | p:vac\rangle  = 0 \quad\text{for each }
  \phi\in\phasesp\commae \nonumber \\
  \langle p:vac | p:vac\rangle  = 1\spce \label{vacuumdef}
\end{gather}
and multiple particle states by
\begin{equation}
  {\hat{\mathrm a}_p[\phi_1]}^\dagger{\hat{\mathrm
  a}_p[\phi_2]}^\dagger\dots| p:vac\rangle \period
\end{equation}
The mapping between the phase space $\phasesp_p$ and the one-particle space
$\quantsp_1$ is given by
\begin{equation}
  \phi \quad\leftrightarrow\quad {\hat{\mathrm a}_p[\phi]}^\dagger |
  p:vac\rangle
\end{equation}
and is really an isomorphism of Hilbert spaces
\begin{gather}
  \langle p:vac| \hat{\mathrm a}_p[\phi_1] {\hat{\mathrm
  a}_p[\phi_2]}^\dagger | p:vac\rangle  =
  \langle p:vac| [ \hat{\mathrm a}_p[\phi_1] , {\hat{\mathrm
  a}_p[\phi_2]}^\dagger ] | p:vac\rangle  =
  \langle \phi_1,\phi_2\rangle_p\commae\\
  i\star\phi \quad\leftrightarrow\quad \hat{\mathrm
  a}_p[i\star\phi]^\dagger\,| p:vac\rangle  = i\hat{\mathrm
  a}_p[\phi]^\dagger\,| p:vac\rangle \period
\end{gather}

We have successfully found a Fock structure in our quantum space $\quantsp$
with one-particle states labeled by vectors from the classical phase space of
the scalar field $\phasesp$. We will call such construction a \emph{particle
interpretation} of the scalar field theory. For this construction we have
used a new element --- the complex structure $J_p$. We can expect that
different complex structures can give us different Fock structures in
$\quantsp$, and we will investigate this question in part
\ref{sec:inoutform}.

We also implicitly assumed that condition (\ref{vacuumdef}) selects a unique
vacuum state (up to a phase) and creation operators acting on the vacuum
state generate a complete set of vectors in $\quantsp$. This assumption
corresponds to the assumption that the set of observables $\qPhi^x$ for $x\in
M$ is a sufficient set of observables for description of our system, i.e.
that we do not have any other degrees of freedom which are not reflected in
field observables $\qPhi^x$. In the opposite case we should use another kind
of field.

 From the strict mathematical point of view we have constructed a particular
representation of the quantum algebra of observables generated by $\qPhi^x$
on a Fock space based on one-particle space isomorphic with $\phasesp_p$.


\subsection*{Positive-negative frequency splitting}

Now we will show a connection with the usual definition of particle states
using a mode expansion of the field operator \cite{BirrellDavies:book}. For
it we have to investigate properties of the complex structure. $J_p$ as an
operator on $\phasesp$ does not have eigenvectors in $\phasesp$, but it has
eigenvectors in the complexification $\phasesp^{\mathbb C} = {\mathbb
C}\otimes \phasesp$ of the phase space. Its eigenvalues are $\pm i$ (because
squares of them have to be $-1$) and we can explicitly write projectors on
subspaces of $\phasesp^{\mathbb C}$ with these eigenvalues
\begin{equation}
  P^\pm_p = \frac12 (\delta\mp i J_p)\period
\end{equation}
They have properties
\begin{gather}
  J_p \circ P^+_p = i P^+_p \comma J_p\circ P^-_p = - i P^-_p
  \commae\nonumber\\
  P^\pm_p\circ P^\pm_p = P^\pm_p \comma P^\pm_p \circ P^\mp_p = 0 \comma
  {P^\pm_p}{}^* = P^\mp_p \commae\label{Psprop}\\
  P^+_p + P^-_p = \delta \comma i (P^+_p - P^-_p) = J_p\period\nonumber
\end{gather}
The compatibility (\ref{JdFcompatibility}) of $J_p$ and $\symplstr$ is
possible to reformulate as
\begin{gather}
  P^\pm_p \circ \symplstr\circ P^\pm_p = 0 \comma
  \symplstr = P^-_p \circ \symplstr\circ P^+_p + P^+_p \circ \symplstr\circ
  P^-_p \period\\
  P^\pm_p \circ \gamma_p\circ P^\pm_p = 0 \comma
  \symplstr = P^-_p \circ \gamma_p\circ P^+_p + P^+_p \circ \gamma_p\circ
  P^-_p \commae\label{PgammaP}\\
  P^\pm_p \circ \GFC\circ P^\pm_p = 0 \comma
  \GFC = P^-_p \circ \GFC \circ P^+_p + P^+_p \circ \GFC\circ P^-_p
  \period\label{GRCposnegparts}
\end{gather}
We will call the \emph{positive} resp. \emph{negative frequency part} of a
solution $\phi$ of the equation of motion (\ref{freeeqofmot}) (i.e.
$\phi\in\phasesp$) the complex solutions $\phi\frqp{+}{p}$ resp.
$\phi\frqp{-}{p}$ of the same equation (i.e.
$\phi\frqp{\pm}{p}\in\phasesp^{\mathbb C}$) defined by
\begin{equation}
  \phi\frqp{\pm}{p} = P^\pm_p \circ \phi\period
\end{equation}
We have
\begin{equation}
  \phi = \phi\frqp{+}{p} + \phi\frqp{-}{p} \comma J_p\circ\phi = i
  (\phi\frqp{+}{p} - \phi\frqp{-}{p})\comma {\phi\frqp{\pm}{p}}{}^* =
  \phi\frqp{\mp}{p}\period
\end{equation}

The scalar product is possible to write as
\begin{equation}
  \langle \phi_1 , \phi_2 \rangle_p =
  {\phi_1}\frqp{-}{p} \circ \gamma_p\circ{\phi_2}\frqp{+}{p}
  = - i\,{\phi_1}\frqp{-}{p} \circ \symplstr\circ{\phi_2}\frqp{+}{p} \period
\end{equation}
We see that the scalar product $\langle \phi_1,\phi_2\rangle_p$ is
essentially the Klein-Gordon product of the negative and positive frequency
parts of $\phi_1$ and $\phi_2$.

Let's note that not all linear operators on $\phasesp$ are also $p$-linear on
$\phasesp_p$. $p$-linearity of an operator $L$ means that $L$ is linear with
respect of multiplication $\star$. Therefore it has to commute with
multiplication by imaginary unity which is given by action of the complex
structure $J_p$, i.e.
\begin{equation}\label{plinearity}
   L\circ J_p = J_p\circ L\period
\end{equation}
Similarly $p$-antilinearity of an operator $A$ is equivalent to
\begin{equation}\label{pantilinerity}
   A\circ J_p = - J_p\circ A\period
\end{equation}

We introduce the hermitian conjugation $L^{\langle\dagger\rangle}$ of a
$p$-linear operator $L$ defined by
\begin{equation}\label{phermconj}
  \langle \phi_1 \circ L^{\langle\dagger\rangle} , \phi_2 \rangle_p
  = \langle \phi_1 ,  L \circ \phi_2 \rangle_p\comma
  L^{\langle\dagger\rangle} = - \GFC \circ L \circ \symplstr \period
\end{equation}
We see that it depends on the choice of $J_p$ only through the $p$-linearity
condition on $L$. Similarly we can define ``transposition''
$A^{\langle\top\rangle}$ for a $p$-antilinear operator $A$
\begin{equation}\label{ptranspos}
  \langle \phi_1 \circ A^{\langle\top\rangle} , \phi_2 \rangle_p^*
  = \langle \phi_1 ,  A \circ \phi_2 \rangle_p\comma
  A^{\langle\top\rangle} = \GFC \circ A \circ \symplstr\period
\end{equation}
Both these operator are particular cases of the transposition $O^{{\mathbf
T}_p}$ of any operator $O$ on $\phasesp$ defined using the bi-form
$\gamma_p$:
\begin{gather}
  O^{{\mathbf T}_p} = \gamma_p^{-1} \circ O \circ
  \gamma_p\commae\label{prealtranspos}\\
  L^{\langle\dagger\rangle} = L^{{\mathbf T}_p} \comma A^{\langle\top\rangle}
  = A^{{\mathbf T}_p}\period
\end{gather}
This operation depends on the choice of $J_p$ explicitly.

We can also express creation and annihilation operators using $P^\pm_p$
\begin{equation}\label{ancrarenegpos}
\begin{split}
  &\hat{\mathrm a}[\phi] = -i\,\phi\frqp{-}{p} \circ \symplstr \circ \qPhi =
  - i\,\phi\circ\symplstr\circ\qPhi\frqp{+}{p}\commae\\
  &\hat{\mathrm a}[\phi]^\dagger = i\,\phi\frqp{+}{p} \circ \symplstr \circ
  \qPhi =
  i\,\phi\circ\symplstr\circ\qPhi\frqp{-}{p}\period
\end{split}
\end{equation}
It means that positive resp. negative frequency part of the field operator
$\qPhi$ is composed only from annihilation resp. creation operators and vice
versa. Using these expressions we can easily prove the commutation relations
(\ref{anihcreatommrel}). For example,
\begin{equation}
\begin{split}
  &\bigl[\hat{\mathrm a}[\phi_1],\hat{\mathrm a}[\phi_2]^\dagger\bigr] =
  -i\,\phi_1\circ
  P^-_p\circ\symplstr\circ\bigl[\qPhi,\qPhi\bigr]\circ\symplstr^{\top}\circ
  P^+_p\circ\phi_2\,i =\\
  &\quad= \phi_1\circ P^-_p \circ\symplstr\circ i \GFC \circ\symplstr\circ
  P^+_p \circ\phi_2\,\opone =
  -i\,\phi_1\circ P^-_p\circ\symplstr\circ P^+_p\circ\phi_2\,\opone=\\
  &\quad= \langle\phi_1,\phi_2\rangle_p\,\opone\period\label{anihcreatommrel}
\end{split}
\end{equation}

Now we can choose a $p$-orthonormal $\mathbb C$-base ${\mathbf u} =
\{u_{\mathrm k}; {\mathrm k}\in {\mathcal I}\}$ in $\phasesp_p$ where
$\mathrm k$ is an index from some, for simplicity discrete, set $\mathcal I$.
$p$-orthonormality means
\begin{equation}
  \langle u_{\mathrm k}, u_{\mathrm l}\rangle_p = \delta_{\mathrm
  {kl}}\period
\end{equation}
${\mathbb C}$-base means that the set ${\mathbf u}$ is complete in the
Hilbert space $\phasesp_p$, i.e. in the vector space with multiplication
$\star$. The set $\{u_{\mathrm k}\frqp{+}{p},u_{\mathrm k}\frqp{-}{p};
{\mathrm k}\in {\mathcal I}\}$ of positive and negative frequency parts of
vectors $u_{\mathrm k}$ forms a complete set in $\phasesp^{\mathbb C}$ with
properties
\begin{gather}
  u_{\mathrm k}\frqp{\pm}{p}\circ\symplstr\circ u_{\mathrm l}\frqp{\pm}{p}=0
  \comma
  -i u_{\mathrm k}\frqp{-}{p}\circ\symplstr\circ u_{\mathrm l}\frqp{+}{p} =
  \delta_{\mathrm {kl}}\commae\nonumber\\
  u_{\mathrm k}\frqp{\pm}{p}{}^* = u_{\mathrm l}\frqp{\mp}{p}\comma F\bullet
  u_{\mathrm k}\frqp{\pm}{p} = 0 \period\label{upmorthonormality}
\end{gather}
These are standard properties \cite{BirrellDavies:book} of modes which are
used for the expansion of a field operator. We can decompose
$\qPhi\in\phasesp\otimes\quantobs$ using the base $\{ u_{\mathrm
k}\frqp{+}{p}, u_{\mathrm k}\frqp{-}{p}; {\mathrm k}\in {\mathcal I}\}$
\begin{equation}
  \qPhi = \sum_{{\mathrm k}\in {\mathcal I}}
  \bigl( \hat{\mathrm a}_{\mathrm k}  u_{\mathrm k}\frqp{+}{p} +
  \hat{\mathrm a}_{\mathrm k}^\dagger u_{\mathrm k}\frqp{-}{p}\bigl)  \commae
\end{equation}
where operator valued coefficients can be found using the relations
(\ref{upmorthonormality})
\begin{equation}\label{cranopforbase}
\begin{split}
  \hat{\mathrm a}_{\mathrm k} &= -i\,{u_{\mathrm k}}\frqp{-}{p} \circ
  \symplstr\circ \qPhi
  = \hat{\mathrm a}_p[u_{\mathrm k}]\commae\\
  \hat{\mathrm a}_{\mathrm k}^\dagger &=
  -i\,\qPhi\circ\symplstr\circ{u_{\mathrm k}}\frqp{+}{p}
  = \hat{\mathrm a}_p[u_{\mathrm k}]^\dagger\period
\end{split}
\end{equation}
We see that the usual mode expansion gives nothing other than our creation
and annihilation operators for a chosen base ${\mathbf u}$ in the phase space
$\phasesp$. This connection also justifies using the words ``positive'' resp.
``negative frequency part'' for vectors from $P^\pm_p \circ \phasesp$.


\subsection*{Green functions}

We can ask whether the construction of the Fock structure in $\quantsp$ using
the complex structure is not artificial. Does $J_p$ have any physical
meaning? Is it connected with any interesting physical quantity? The answer
is yes. The complex structure has a rather simple interpretation. To show it
we have to introduce Green functions associated with a particular choice of
vacuum state.

The Wightman functions are defined by
\begin{equation}
  \GFP_p{}^{xy} = \langle p:vac|\qPhi^x\qPhi^y| p:vac\rangle \comma
  \GFM_p{}^{xy} = \langle p:vac|\qPhi^y\qPhi^x| p:vac\rangle \period
\end{equation}
The Hadamard Green function is
\begin{equation}
  \GFH_p{}^{xy} = \langle p:vac| \qPhi^x\qPhi^y + \qPhi^y\qPhi^x |
  p:vac\rangle \period
\end{equation}
The causal Green function can be written as
\begin{equation}
  -i \GFC^{xy} = \langle p:vac|\qPhi^x\qPhi^y - \qPhi^y\qPhi^x| p:vac\rangle
  \period
\end{equation}
All these Green functions are solution of the equation of motion
(\ref{freeeqofmot}) in both arguments and therefore they are bi-vectors from
$\phasesp^2_0$.

We can also define the Feynman Green function
\begin{equation}
  \GFF_p{}^{xy} = \langle p:vac| {\mathcal T}\bigl(\qPhi^x\qPhi^y\bigr)|
  p:vac\rangle   \commae
\end{equation}
where ${\mathcal T}$ denotes time ordering of operators
\begin{equation}
  {\mathcal T}\bigl(\qPhi^x\qPhi^y\bigr) =
  \begin{cases}
  \qPhi^x\qPhi^y &\text{for $x$ after $y$}\\
  \qPhi^y\qPhi^x &\text{for $y$ after $x$}\period
  \end{cases}
\end{equation}
For $x,y$ space-like separated $\qPhi^x$ and $\qPhi^y$ commute, so the
definition is unique. Finally we define retarded, advanced and symmetric
Green functions\note{nonhomGFeq}
\begin{gather}
  (\chi[\domain]\,\delta) \bullet F \bullet \GFRA = - (\chi[\domain]\,\delta)
  \quad \text{for any}\quad \domain =
  \langle\Sigma_f|\Sigma_i\rangle\commae\nonumber\\
  \GFR^{xy} = 0 \quad \text{for $x$ before $y$}\comma
  \GFA^{xy} = 0 \quad \text{for $x$ after $y$}\commae
\end{gather}
\begin{equation}
  \GFS = \frac12 (\GFR + \GFA)\period
\end{equation}

Relations among these Green functions are
\begin{gather}
  \GFH_p = \GFP_p + \GFM_p\comma
  \GFC = i (\GFP_p - \GFM_p)\commae\label{relGFHCtoPM}\\
  \GFP_p = \frac12(\GFH_p -i\GFC)\comma
  \GFM_p = \frac12(\GFH_p +i\GFC)\commae \label{relGFPMtoHC}\\
  \GFP_p{}^\top = \GFM_p\comma \GFH_p{}^\top = \GFH_p\comma
  \GFF_p{}^\top = \GFF_p\comma \GFC{}^\top = - \GFC\commae \label{GFtransp}\\
  \GFP_p{}^* = \GFM_p\comma \GFH_p{}^* = \GFH_p\comma
  \GFF_p{}^* = \GFF_p\comma \GFC{}^* = \GFC\commae \label{GFconjug}\\
  \GFC = \GFR - \GFA\comma
  \GFS = \frac12(\GFR+\GFA)\commae
\end{gather}
\begin{equation}\label{relGFFtoMPH}
\GFF_p{}^{xy} =
\begin{cases}
  \GFP_p{}^{xy}&\text{for $x$ after $y$}\commae\\
  \GFM_p{}^{xy}&\text{for $x$ befor $y$}\commae\\
  \GFP_p{}^{xy} = \GFM_p{}^{xy} = \frac12 \GFH_p{}^{xy}&\text{for $x, y$
  space-like separated}\period
\end{cases}
\end{equation}
Using these relations we can derive
\begin{equation}\label{relGFFtoSH}
  \GFF_p = - i \GFS + \frac12\GFH_p
\end{equation}
and therefore the Feynman Green function satisfies\note{nonhomGFeq}
\begin{equation}\label{eqforGFF}
  -i (\chi[\domain]\,\delta) \bullet F \bullet\GFF_p =
  (\chi[\domain]\,\delta)
  \quad \text{for any}\quad \domain = \langle\Sigma_f|\Sigma_i\rangle\period
\end{equation}
This means that Green functions $\GFF_p, \GFR, \GFA, \GFS$ are not tensors
from $\phasesp^2_0$ but rather they belong to $\hist^2_0$.

Let's note that Green functions $\GFR, \GFA, \GFS$ and $\GFC$ are independent
on the choice of the complex structure $J_p$. Green functions $\GFP_p,
\GFM_p, \GFH_p$ and $\GFF_p$ depend of the choice of vacuum state and
therefore on the choice of  the particle interpretation.

After introducing this whole zoo of Green functions we will pay more
attention to $\GFP_p, \GFM_p$ and $\GFH_p$. Using the fact that the positive
frequency part of $\qPhi$ annihilates the vacuum state we get
\begin{equation}\label{PGFPMP}
  \GFP_p = P^+_p \circ \GFP_p \circ P^-_p\comma
  \GFM_p  = P^-_p \circ \GFP_p \circ P^+_p\period
\end{equation}
Using equations (\ref{relGFHCtoPM}), (\ref{GRCposnegparts}) and
(\ref{Psprop}) we find
\begin{gather}
  \GFPM_p = \mp i\,P^\pm_p \circ \GFC \circ P^\mp_p = \mp i\,P^\pm_p \circ
  \GFC\commae\\
  \GFH_p = -i\,(P^+_p - P^-_p) \circ \GFC = - J_p \circ \GFC =
  {\gamma_p}^{-1}\spce
\end{gather}
or
\begin{gather}
  P^\pm_p = \mp i\,\GFPM_p \circ \symplstr\commae \label{relPPMGFPM}\\
  J_p = \GFH_p \circ \symplstr\period \label{relJGFH}
\end{gather}
This means that the complex structure $J_p$ is essentially the Hadamard Green
function. Or, more precisely, the action of $J_p$ on a solution
$\phi\in\phasesp$ is given by Klein-Gordon product of the Hadamard Green
function with the solution $\phi$. Wightman functions are in similar
relations with projector operators $P^\pm_p$.

We can use this relation to derive compositions laws for Green functions. The
translation of equations (\ref{Psprop}), (\ref{Jsquare}) to the language of
Green functions gives us
\begin{gather}
  \GFPM_p \circ\symplstr\circ\GFPM_p = \pm i\,\GFPM_p\comma\GFPM_p
  \circ\symplstr\circ\GFMP_p = 0\commae\\
  \GFH_p \circ\symplstr\circ\GFH_p = \GFC\period
\end{gather}
Using (\ref{relGFFtoMPH}) we get
\begin{equation}
  (\chi[\domain_f]\,\delta)\bullet
  \GFF_p\bullet\symplstr[\Sigma]\bullet\GFF_p\bullet
  (\chi[\domain_i]\,\delta) =
  i\,(\chi[\domain_f]\,\delta)\bullet\GFF_p\bullet (\chi[\domain_i]\,\delta)
\end{equation}
for any Cauchy hypersurface $\Sigma$ and spacetime domains $\domain_f$ resp.
$\domain_i$ in the future resp. the past of the hypersurface $\Sigma$.


\subsection*{Total particle-number observable}

We can define a quantum observable of the number of particles in a state
labeled by a solution $\phi$
\begin{equation}
  \hat{\mathrm n}_p[\phi] = \frac{\hat{\mathrm a}_p[\phi]^\dagger
  \hat{\mathrm a}_p[\phi]}{\langle\phi,\phi\rangle_p}\period
\end{equation}
It satisfies
\begin{equation}
  \hat{\mathrm n}_p[\phi]\,\hat{\mathrm a}_p[\phi]^{\dagger\,m} |
  p:vac\rangle  = m\,\hat{\mathrm a}_p[\phi]^{\dagger\,m} | p:vac\rangle
  \period
\end{equation}

Let's choose again a $p$-orthonormal ${\mathbb C}$-base ${\mathbf u} =
\{u_{\mathrm k}; {\mathrm k}\in {\mathcal I}\}$ in $\phasesp_p$. It generates
an orthonormal base in quantum space $\quantsp$ composed of particle states
\begin{equation}\label{partbase}
  |p\,{\mathbf u}:\boldsymbol{m}\rangle = \frac1{\sqrt{\boldsymbol{m}!}}
  \hat{\mathrm a}_p[{\mathbf u}]^{\dagger\,\boldsymbol{m}} |
  p:vac\rangle\period
\end{equation}
Here $\boldsymbol{m} = \{m_{\mathrm k}; {\mathrm k}\in {\mathcal I} \}$ is a
multiindex and we are using shorthands
\begin{equation}\label{shorthands}
  \boldsymbol{m}! = \prod_{{\mathrm k}\in {\mathcal I}} m_{\mathrm k}! \comma
  \hat{\mathrm a}[{\mathbf u}]^{\dagger\,\boldsymbol{m}}
  =  \prod_{{\mathrm k}\in {\mathcal I}} \hat{\mathrm a}_p[u_{\mathrm
  k}]^{\dagger\,m_{\mathrm k}}
  =  \prod_{{\mathrm k}\in {\mathcal I}} \hat{\mathrm a}_{\mathrm
  k}^{\dagger\,m_{\mathrm k}}  \commae
\end{equation}
where creation operators $\hat{\mathrm a}_{\mathrm k}$ are defined by Eq.\
(\ref{cranopforbase}). The combinatoric factor is chosen so that states are
normalized:
\begin{equation}
  \langle p\,{\mathbf u}:\boldsymbol{m}|p\,{\mathbf u}:\boldsymbol{m'}\rangle
  = \delta_{\boldsymbol{mm'}}\period
\end{equation}

The observable of the number of particles in a mode $u_{\mathrm k}$ is
\begin{equation}\label{defofnk}
  \hat{\mathrm n}_{p\,{\mathrm k}} = \hat{\mathrm n}_p[u_{\mathrm k}]
  = \hat{\mathrm a}_{\mathrm k}^\dagger \hat{\mathrm a}_{\mathrm k} \commae
\end{equation}
and it satisfies
\begin{equation}
  \hat{\mathrm n}_{p\,{\mathrm k}} |p\,{\mathbf u}:\boldsymbol{m}\rangle =
  m_{\mathrm k} |p\,{\mathbf u}:\boldsymbol{m}\rangle\period
\end{equation}

Now we can define the observable of the total number of $p$-particles
\begin{equation}
  \hat{\mathrm N}_p = \sum_{{\mathrm k}\in {\mathcal I}} \hat{\mathrm
  n}_{p\,{\mathrm k}}\period
\end{equation}
Using definitions (\ref{cranopforbase}) of creation and annihilation
operators and orthonormality of the base ${\mathbf u}$ we get
\begin{equation}
  \hat{\mathrm N}_p =
  \sum_{{\mathrm k}\in {\mathcal I}}
  \hat{\mathrm a}_p[u_{\mathrm k}]^\dagger \hat{\mathrm a}_p[u_{\mathrm k}] =
  \sum_{{\mathrm k}\in {\mathcal I}}
  \langle \qPhi, u_{\mathrm k}\rangle_p \langle u_{\mathrm k}, \qPhi
  \rangle_p =
  \langle \qPhi, \qPhi \rangle_p\period
\end{equation}
We see that $\hat{\mathrm N}_p$ is independent on the choice of the base
${\mathbf u}$ but it depends on the complex structure $J_p$ through the
scalar product and therefore it depends on the particle interpretation. It
will be useful to write down also another representation of $\hat{\mathrm
N}_p$. First let's note
\begin{equation}
  \qPhi\circ\symplstr\circ\qPhi =
  - i \bigl(\frac12 \tr_{\phasesp} \delta^{(\phasesp)}\bigr)\,\opone =
  - i \bigl( \tr_{\phasesp_p} \delta^{(\phasesp_p)}\bigr)\,\opone \period
\end{equation}
So
\begin{equation}
  \hat{\mathrm N}_p = \langle \qPhi, \qPhi \rangle_p = \frac12
  \qPhi\circ(\gamma_p - i \symplstr)\circ\qPhi =
  \frac12
  \qPhi\circ\gamma_p\circ\qPhi-\frac12\bigl(\tr_{\phasesp_p}\delta^{(%
\phasesp_p)}\bigr)\,\opone\period
\end{equation}

We see that $\hat{\mathrm N}_p$ is a quantum version of a classical quadratic
observable
\begin{equation}
  N_p = \frac12 \Phi\circ\gamma_p\circ\Phi
\end{equation}
with special operator ordering. This operator ordering is called $p$-normal
ordering and is defined by the condition that in any product of operators all
$p$-creation operators are on the left of all $p$-annihilation operators. We
denote a quantum version of a classical observable $F(\Phi)$ in $p$-normal
ordering by
\begin{equation}
  \hat{\mathrm F} = \nordl F(\qPhi) \nordr_p\period
\end{equation}
It can be written more explicitly for any quadratic observable defined using
a symmetric bi-form $k\in\phasesp^0_2$
\begin{equation}
\begin{split}
  K(\Phi) &= \frac12 \Phi\circ k\circ\Phi
  = \frac12 (\Phi\frqp{+}{p} + \Phi\frqp{-}{p})\circ k\circ(\Phi\frqp{+}{p} +
  \Phi\frqp{-}{p}) = \\
  &= \frac12 \Phi\frqp{+}{p} \circ k \circ \Phi\frqp{+}{p} + \frac12
  \Phi\frqp{-}{p} \circ k \circ \Phi\frqp{-}{p} +
    \Phi\frqp{-}{p} \circ k \circ \Phi\frqp{+}{p}\commae
\end{split}
\end{equation}
\begin{equation}
  \hat{\mathrm K} = \nordl  K(\qPhi) \nordr_p =
  \frac12 \qPhi\frqp{+}{p} \circ k \circ \qPhi\frqp{+}{p} + \frac12
  \qPhi\frqp{-}{p} \circ k \circ \qPhi\frqp{-}{p} +
  \qPhi\frqp{-}{p} \circ k \circ \qPhi\frqp{+}{p}\period
\end{equation}
For the observable $N_p(\Phi)$ using Eq.\ (\ref{PgammaP}) we get
\begin{equation}
  \hat{\mathrm N}_p = \qPhi\frqp{-}{p} \circ \gamma_p \circ \qPhi\frqp{+}{p}
  = \nordl  N_p(\qPhi) \nordr_p\period
\end{equation}


\subsection*{Diagonalization of Hamiltonian}

Let's assume that a classical quadratic positive definite Hamiltonian is
given,
\begin{equation}
  H(\Phi) = \frac12 \Phi\circ h\circ\Phi  \commae
\end{equation}
where $h$ is a positive definite symmetric bi-form from $\phasesp^0_2$. We
can ask whether there exists a particle interpretation such that particle
states have definite energy and the energy is additive with respect to the
number of particles. More precisely, we will look for such a complex
structure $J_p$  and $p$-orthonormal base ${\mathbf u}=\{u_{\mathrm k};
{\mathrm k}\in {\mathcal I}\}$ which satisfy
\begin{equation}
  \hat{\mathrm H}\,|p\,{\mathbf u}:\boldsymbol{m}\rangle =
  \Bigl(\sum_{{\mathrm k}\in {\mathcal I}} \omega_{\mathrm k} m_{\mathrm
  k}\Bigr)\,|p\,{\mathbf u}:\boldsymbol{m}\rangle  \commae
\end{equation}
where $\omega_{\mathrm k}\in {\mathbb R}^+$ is the energy of the one-particle
state $\hat{\mathrm a}_{\mathrm k}^\dagger\,| p:vac\rangle $
\begin{equation}
  \hat{\mathrm H}\,\hat{\mathrm a}_{\mathrm k}^\dagger\,| p:vac\rangle  =
  \omega_{\mathrm k}\,\hat{\mathrm a}_{\mathrm k}^\dagger\,| p:vac\rangle
  \period
\end{equation}
This requirement is equivalent to the requirement that the Hamiltonian have
the form
\begin{equation}
  \hat{\mathrm H} = \sum_{{\mathrm k}\in {\mathcal I}} \omega_{\mathrm k}
  \hat{\mathrm n}_{p\,{\mathrm k}}\period
\end{equation}
Using the definition of $\hat{\mathrm n}_{p\,{\mathrm k}}$ (\ref{defofnk})
and the orthonormality of the base ${\mathbf u}$, we get
\begin{equation}\label{HusingOm}
  \hat{\mathrm H} = \sum_{{\mathrm k}\in {\mathcal I}} \langle \qPhi,
  u_{\mathrm k}\rangle_p \omega_{\mathrm k}
  \langle  u_{\mathrm k}, \qPhi \rangle_p = \langle \qPhi, \Omega
  \circ\qPhi\rangle_p
  = \nordl  \qPhi \circ \gamma_p\circ\Omega\circ\qPhi \nordr_p  \commae
\end{equation}
where $\Omega$ is a $p$-linear hermitian positive definite operator on
$\phasesp_p$ given by
\begin{equation}
  \Omega\circ u_{\mathrm k} = \omega_{\mathrm k}\,u_{\mathrm k}\period
\end{equation}
$p$-linearity gives us the condition
\begin{equation}\label{Omlinearity}
  \bigl[ \Omega , J_p \bigr] = 0\period
\end{equation}
Hermicity $\Omega^{\langle\dagger\rangle} = \Omega$ (following from
$\omega_{\mathrm k}\in {\mathbb R}^+$) gives
\begin{equation}\label{Omhermicity}
   \symplstr\circ\Omega = \Omega\circ\symplstr \quad\Leftrightarrow\quad
   \gamma_p\circ\Omega = \Omega\circ\gamma_p\period
\end{equation}

The quantum observable $\hat{\mathrm H}$ should be a quantum version of the
classical observable $H(\Phi)$ in some operator ordering. As we have
discussed, any two quantum versions of $H(\Phi)$ defined using two different
operator orderings can differ only by multiple of the unit operator.
Therefore we can write
\begin{equation}
  \hat{\mathrm H} = \nordl  H(\qPhi) \nordr_p + \alpha \opone
  = \nordl\qPhi{} \circ h\circ \qPhi{} \nordr_p + \alpha\opone\period
\end{equation}
 From a comparison with Eq.\ (\ref{HusingOm}) we see that we need to satisfy
\begin{equation}\label{hJOmrel}
  h = \gamma_p \circ \Omega \quad\Leftrightarrow\quad \GFC\circ h =
  J_p\circ\Omega\period
\end{equation}

Let's summarize. We are looking for a complex structure $J_p$ and a positive
definite operator $\Omega$ which satisfy conditions (\ref{JdFcompatibility}),
(\ref{JdFpositivity}), (\ref{Omlinearity}), (\ref{Omhermicity}) and
(\ref{hJOmrel}). We can get such a $J_p$ and $\Omega$ using a polar
decomposition\note{polardecomp} of the operator $(\GFC\circ h)$. For a polar
decomposition we need a transposition of operators. Let us use the
transposition defined using a positive definite symmetric bi-form $h$
\begin{equation}
  A^{\mathbf T} = h^{-1}\circ A \circ h \qquad \text{for any operator $A$ on
  $\phasesp$}\period
\end{equation}
Because $\GFC\circ h$ is antisymmetric with respect of this transposition,
left and right polar decompositions of $(\GFC\circ h)$ coincide and we can
define
\begin{align}
  \Omega &= \abs{ \GFC\circ h } = \bigl( (\GFC\circ h)^{\mathbf T} \circ
  (\GFC\circ h)\bigr)^{\frac12}
  = \bigl( (\GFC\circ h) \circ (\GFC\circ h)^{\mathbf
  T}\bigr)^{\frac12}\commae \label{Omegausingh}\\
  J_p &= \sign\bigl(\GFC\circ h\bigr) = (\GFC\circ h)\circ\Omega^{-1}
  = \Omega^{-1}\circ(\GFC\circ h) \commae\label{Jusingh}\\
  \GFC \circ h &= J_p\circ\Omega = \Omega\circ J_p\period \label{hJOmrel2}
\end{align}
It is straightforward to check all conditions on $J_p$ and $\Omega$. Positive
definiteness and symmetry ($\Omega^{\mathbf T} = \Omega$) of $\Omega$ follows
from the definition of square root, the compatibility of $J_p$ and
$\symplstr$ follows from Eq.\ (\ref{Jusingh}) and symmetry of $\Omega$,
positive definiteness of $\gamma_p$ follows from Eq.\ (\ref{Jusingh}) and
positive definiteness of $\Omega$, (\ref{Omlinearity}) and (\ref{hJOmrel})
are the same as Eq.\ (\ref{hJOmrel2}).

We have finally proved that the positive quadratic classical Hamiltonian
$H(\Phi)$ picks up uniquely the particle interpretation in which it is
possible to diagonalize the quantum Hamiltonian, i.e. to write
\begin{equation}
  \hat{\mathrm H} = \nordl H(\qPhi) \nordr_p =
  \sum_{{\mathrm k}\in {\mathcal I}} \omega_{\mathrm k} \hat{\mathrm
  n}_{p\,{\mathrm k}}  \commae
\end{equation}
where $\hat{\mathrm n}_{p\,{\mathrm k}}$ are observables of number of
particle in modes $u_{\mathrm k}$ which are eigenvectors of the operator
$\Omega$ with eigenvalues $\omega_{\mathrm k}$. A similar result is possible
to find in \cite{AshtekarMagnon:1975,Gribetal:book}.

Unfortunately we do not have a preferable 3+1 splitting in a general
spacetime, and for a general 3+1 splitting the Hamiltonian is time dependent.
This means that the diagonalization criterion picks up different particle
interpretations at different times. This reflects the fact that in a general
spacetime we do not have a preferable particle interpretation and if we
decide to choose particle interpretations connected with the Hamiltonian of
some 3+1 decomposition, we have to expect particle creation as it will be
described in part \ref{sec:inoutform}.


\subsection*{Boundary conditions}

We have seen that Green functions $\GFPM_p, \GFH_p, \GFF_p$ are associated
with each particle interpretation. The Green functions of the same kind for
different particle interpretations satisfy the same equations and symmetry
conditions. They differ by boundary conditions. Now we will find what
boundary conditions are associated with a particle interpretation.

First we will reformulate structures we have introduced earlier in the
language of quantities localized on a hypersurface $\Sigma$. Space field
quantities are tensors from subspaces of $\hist^k_l$ or $\phasesp^k_l$
defined using the projector operator $D_\varphi[\Sigma]$. Therefore we need
to express quantities as $J_p, P^\pm_p$ using objects from these subspaces.
But a general tensor from $\hist^k_l$ or from $\phasesp^k_l$ has also
components in subspaces which are generated by the projector $D_\pi[\Sigma]$.
However we can use the bi-forms $\distdFl[\Sigma], \distdFr[\Sigma]$ to map
these subspaces to the subspace given by $D_\varphi[\Sigma]$.

We can decompose the complex structure $J_p$ as
\begin{equation}
  J_p = \AoperH_p - \BmetrH_p\circ\distdFr + \distdFr^{-1}\circ \CmetrH_p
  + \distdFr^{-1}\circ \DoperH_p\circ \distdFr  \commae
\end{equation}
where
\begin{gather}
  \AoperH_p, \DoperH_p\in \phasesp^1_1 \comma \BmetrH_p\in\phasesp^2_0\comma
  \CmetrH_p\in\phasesp^0_2\commae\\
\begin{split}
  \AoperH_p\circ D_\pi = D_\pi \circ \AoperH_p = 0&\comma
  \DoperH_p\circ D_\pi = D_\pi \circ \DoperH_p = 0\commae\\
  \BmetrH_p\circ D_\pi = D_\pi \circ \BmetrH_p = 0&\comma
  \CmetrH_p\circ D_\pi = D_\pi \circ \CmetrH_p = 0\period
\end{split}
\end{gather}
For brevity of notation we drop the explicit dependence on $\Sigma$.

The conditions (\ref{Jsquare}), (\ref{JdFcompatibility}) give
\begin{gather}
  \DoperH_p = - \AoperH_p\comma
  \CmetrH_p = \CmetrH_p^\top\comma \BmetrH_p =
  \BmetrH_p^\top\commae\nonumber\\
  \AoperH_p\circ \BmetrH_p = \BmetrH_p\circ \AoperH_p\comma
  \AoperH_p\circ \CmetrH_p = \CmetrH_p\circ \AoperH_p\commae\\
  \BmetrH_p\circ \CmetrH_p = \CmetrH_p \circ \BmetrH_p = D_\varphi +
  \AoperH_p\circ \AoperH_p\period\nonumber
\end{gather}
So we can write
\begin{gather}
  J_p = \AoperH_p - \BmetrH_p\circ\distdFr + \distdFr^{-1}\circ \CmetrH_p
  - \distdFr^{-1}\circ \AoperH_p\circ \distdFr\commae\label{JusingABC}\\
  \gamma_p = \CmetrH_p - \distdFl\circ \AoperH_p - \AoperH_p\circ \distdFr
  + \distdFl\circ \BmetrH_p \circ\distdFr\commae\label{gammausingABC}\\
  \GFH_p = \BmetrH_p + \distdFr^{-1}\circ \AoperH_p + \AoperH_p\circ
  \distdFl^{-1}
  + \distdFr^{-1}\circ \CmetrH_p
  \circ\distdFl^{-1}\commae\label{GFHusingABC}\\
  P^+_p = \frac12 (D_\varphi + i \distdFr^{-1}\circ\Theta_p)^* \circ
  \BmetrH_p\circ (\Theta_p + i \distdFr)\commae\label{PusingABC}\\
  \frac12 (J_p\circ\symplstr - i\,\symplstr) =
  (\Theta_p + i \distdFl)^* \circ \BmetrH_p\circ(\Theta_p + i
  \distdFr)\commae\label{scprodusingABC}
\end{gather}
where
\begin{equation}
\begin{split}
  \Theta_p = \BmetrH_p^{-1}\circ(D_\varphi - i \AoperH_p) = C\circ(D_\varphi
  + i A)^{-1}\commae\\
  \Theta_p^\top = \Theta_p\comma \Theta_p\circ D_\pi = D_\pi\circ \Theta_p =
  0\period
\end{split}
\end{equation}

Using Eq.\ (\ref{PusingABC}), we find conditions for positive resp. negative
frequency solutions in the language of quantities on $\Sigma$ only:
\begin{equation}\label{PMpartsboundcond}
  (\Theta_p + i \distdFr)\circ\phi\frqp{-}{p} = 0\comma
  (\Theta_p^* - i \distdFr)\circ\phi\frqp{+}{p} = 0\period
\end{equation}
We see that the value and normal derivative on $\Sigma$ of positive resp.
negative frequency solutions are ``proportional'' to each other through the
bi-form $\Theta_p$ which is uniquely given by $J_p$ and the choice of the
hypersurface $\Sigma$. It means that the value of the field on a hypersurface
$\Sigma$ is enough to determine the positive resp. negative frequency
solution.

Remembering properties (\ref{PGFPMP}) and (\ref{GFtransp}), (\ref{GFconjug})
we can formulate conditions for $\GFPM_p$
\begin{equation}
\begin{split}
  (\Theta_p^* - i \distdFr)\circ\GFP_p = 0 &\comma \GFP_p \circ (\Theta_p + i
  \distdFl) = 0\commae\\
  (\Theta_p + i \distdFr)\circ\GFM_p = 0 &\comma \GFM_p \circ (\Theta_p^* - i
  \distdFl) = 0\period
\end{split}
\end{equation}
Here it does not matter on which hypersurface we formulate boundary
conditions. Using relations (\ref{relGFFtoMPH}) we get a set of conditions
which uniquely determine the Feynman Green function
\begin{gather}
  -i\, (\chi[\domain]\delta) \bullet F \bullet \GFF_p =
  (\chi[\domain]\,\delta) \commae\qquad\nonumber\\
  (\Theta_p[\Sigma_f]^* - i \distdFr[\Sigma_f]) \bullet \GFF_p \bullet
  (\chi[\domain]\delta)   = 0 \commae\nonumber\\
  (\chi[\domain]\delta) \bullet\GFF_p\bullet (\Theta_p[\Sigma_f]^* - i
  \distdFl[\Sigma_f]) = 0 \commae \label{GFFboundcond}\\
  (\Theta_p[\Sigma_i] + i \distdFr[\Sigma_i]) \bullet \GFF_p \bullet
  (\chi[\domain]\delta) = 0 \commae\nonumber\\
  (\chi[\domain]\delta) \bullet\GFF_p\bullet (\Theta_p[\Sigma_i] + i
  \distdFl[\Sigma_i]) = 0 \spce\nonumber
\end{gather}
for any domain $\domain = \langle\Sigma_f|\Sigma_i\rangle$ and we recovered
explicit dependence on hypersurfaces which is important for boundary
conditions for the Feynman Green function.


\section{In-out formalism}
\label{sec:inoutform}


\subsection*{Two particle interpretations}

Until now we have investigated a single particle interpretation of the scalar
field theory. But there exist a lot of different particle interpretations,
each corresponding to a different complex structure on the phase space, and
in a general situation none of them have a preferred position.

However in most physical situations, we are dealing with spacetime which has
special properties at least in the remote past and future. For example, the
spacetime may be static in these regions. It gives us the possibility to
choose a preferred notion of particles in the past and in the future. These
particle interpretations are not generally the same. It is usual to call one
of these notions of particles ``in'' (or initial) particles and other one
``out'' (or final) particles. We have to face a natural physical question of
what is the relation of these two kinds of particles.

Therefore we need to investigate a relation of two particle interpretation.
This problem is usually described in terms of Bogoljubov coefficients. We
will reformulate the theory using quantities independent on a choice of a
bases of modes, and we will also find connections among different in-out
Green functions and their geometrical interpretation similar to the
interpretation of Green functions associated with the one-particle
interpretation. We will also find that given a Green function with certain
properties, we are able to reconstruct two particle interpretations for which
the Green function is an in-out Green function.


\subsection*{Green functions}

Let us choose two particle interpretations given by two complex structures
$J_i$ and $J_f$. We will change the letter ``$p$'' to the letters ``$i$''
resp. ``$f$'' in all quantities defined in the part \ref{sec:partint}. It
means that we have two, generally different, vacuum states $|i:vac\rangle,
|f:vac\rangle$, two sets of creation and annihilation operators etc..

We can define new Green functions, beside $\GFH_i$, $\GFPM_i$, $\GFF_i$,
$\GFH_f$, $\GFPM_f$, $\GFF_f$. Let's define in-out Hadamard Green function
\begin{equation} \label{fiGFHdef}
  \GFH{}^{xy} = \frac{\langle f:vac|\qPhi^x\qPhi^y +
  \qPhi^y\qPhi^x|i:vac\rangle}{\langle f:vac|i:vac\rangle}\commae
\end{equation}
Wightman functions
\begin{equation}
  \GFP{}^{xy} = \frac{\langle f:vac|\qPhi^x\qPhi^y|i:vac\rangle}{\langle
  f:vac|i:vac\rangle}\comma
  \GFM{}^{xy} = \frac{\langle f:vac|\qPhi^y\qPhi^x|i:vac\rangle}{\langle
  f:vac|i:vac\rangle}\spce
\end{equation}
and Feynman Green function
\begin{equation}
  \GFF{}^{xy} = \frac{\langle f:vac|{\mathcal
  T}\bigl(\qPhi^x\qPhi^y\bigr)|i:vac\rangle}{\langle
  f:vac|i:vac\rangle}\period
\end{equation}
Similar relations to Eqs.\ (\ref{relGFHCtoPM}-\ref{relGFFtoSH}) hold among
these Green functions except relations (\ref{GFconjug}), i.e.
\begin{gather}
  \GFH = \GFP + \GFM\comma\GFC = i (\GFP -
  \GFM)\commae\label{oirelGFHCtoPM}\\
  \GFP = \frac12(\GFH -i\GFC)\comma\GFM = \frac12(\GFH +i\GFC)\commae
  \label{oirelGFPMtoHC}\\
  \GFP{}^\top = \GFM\comma \GFH{}^\top = \GFH\comma \GFF{}^\top = \GFF\comma
  \GFC{}^\top = - \GFC\commae \label{oiGFtransp}\\
  \GFF = - i \GFS + \frac12\GFH\commae\label{firelGFFtoSH}
\end{gather}
\begin{equation}\label{oirelGFFtoMPH}
  \GFF{}^{xy} =
  \begin{cases}
  \GFP{}^{xy}&\text{for $x$ after $y$}\commae\\
  \GFM{}^{xy}&\text{for $x$ befor $y$}\commae\\
  \GFP{}^{xy} = \GFM{}^{xy} = \frac12 \GFH{}^{xy}&\text{for $x, y$ space-like
  separated}\commae
  \end{cases}
\end{equation}
\begin{gather}
  F\bullet\GFH = 0\comma F\bullet\GFPM = 0\commae\nonumber\\
  -i(\chi[\domain]\,\delta)\bullet F\bullet\GFF =
  (\chi[\domain]\,\delta)\period
\end{gather}

Let us define operators $J$ and $P^\pm$ on $\phasesp^{\mathbb C}$ using
equations similar to Eqs.\ (\ref{relPPMGFPM}), (\ref{relJGFH})
\begin{gather}
  P^\pm = \mp i\,\GFPM\circ\symplstr\commae \label{oirelPPMGFPM}\\
  J = \GFH\circ\symplstr = i\,(P^+ - P^-) \period \label{oirelJGFH}
\end{gather}
Using the fact that a positive frequency part of $\qPhi$ annihilates a vacuum
we get
\begin{equation}\label{GFPMPiPfrel}
  \GFP = P^+_f\circ\GFP = \GFP\circ P^-_i\comma
  \GFM = P^-_i\circ\GFM = \GFM\circ P^+_f\commae
\end{equation}
therefore
\begin{equation}
\begin{split} \label{relPifP}
  &P^+ = P^+_f\circ P^+ = P^+ \circ P^+_i \comma
  0 = P^-_f\circ P^+ = P^+ \circ P^-_i \commae\\
  &P^- = P^-_i\circ P^- = P^- \circ P^-_f \comma
  0 = P^+_i\circ P^- = P^- \circ P^+_f \spce
\end{split}
\end{equation}
and finally
\begin{equation}
  P^\pm\circ P^\mp = 0\comma \GFPM\circ\symplstr\circ\GFMP = 0\period
\end{equation}
 From Eq.\ (\ref{oirelGFHCtoPM}) we get
\begin{equation}\label{ortogofPs}
  P^+ + P^- = \delta
\end{equation}
and it, together with previous equation, gives
\begin{gather}
  P^\pm\circ P^\pm = P^\pm \comma \GFPM\circ\symplstr\circ\GFPM = \pm
  i\,\GFPM\commae\label{fiPPisP}\\
  J\circ J = - \delta \comma \GFH\circ\symplstr\circ\GFH = \GFC\period
  \label{fiJJismdelta}
\end{gather}
We can further show
\begin{equation}
  P^\pm \circ\symplstr\circ P^\pm = 0 \comma
  J\circ \symplstr = - \symplstr\circ J \period \label{fiJdFcompatibility}
\end{equation}

We see that $J$ is a complex structure compatible with $\symplstr$, and
$P^\pm$ are its eigenspace projectors. However, there is a difference from
the complex structures $J_i$ or $J_f$ --- the complex structure $J$ does not
act on the space $\phasesp$ but on the complexified space $\phasesp^{\mathbb
C}$. More precisely, in general $J^*\not=J$, $\GFH{}^*\not=\GFH$,
$P^\pm{}^*\not=P^\mp$, $\GFPM{}^*\not=\GFMP$. Therefore the complex structure
$J$ does not define a particle interpretation (except, of course, in the
degenerate case $J_f=J_i=J$).

We can define real and imaginary parts of the complex structure $J$
\begin{gather}
  J = M + i N\period\\
  M = \re J = \frac12 (J+J^*)\comma N = \im J = -i\frac12 (J-J^*)\commae
  \label{MNdef}
\end{gather}
Using Eqs.\ (\ref{ortogofPs}), (\ref{relPifP}) and their complex conjugates
and previous definitions we get
\begin{equation}\label{MNusingPs}
\begin{split}
  M &= i\,(P^+ - P^{-*}) = i\,(P^{-*}-P^-)\commae\\
  N &= P^+ + P^{+*} - \delta = \delta - P^- - P^{-*}\spce
\end{split}
\end{equation}
and
\begin{equation}
  - J_f\circ M = - M\circ J_i = \delta + N \comma
  - M\circ J_f = - J_i\circ M = \delta - N \period \label{NusingMJs}
\end{equation}

This gives us an important relation
\begin{equation} \label{MusingJs}
  \bigl[ -\frac12 ( J_i + J_f) \bigr]\circ M = M \circ \bigl[ -\frac12 ( J_i
  + J_f) \bigr] = \delta
\end{equation}
or
\begin{equation}
  \frac12 (\GFH + \GFH{}^*) = \bigl[\frac12 (\gamma_i + \gamma_f)\bigr]^{-1}
  = -\,\GFC\circ \bigl[\frac12(\GFH_i + \GFH_f)\bigr]^{-1}\circ\GFC\period
\end{equation}
We will define a real symmetric bi-form $\gamma$
\begin{equation}
  \gamma = \bigl[\re\GFH\bigr]^{-1} = \frac12 (\gamma_i + \gamma_f)\period
\end{equation}
This means that $\gamma$, and therefore also $\re\GFH$, are positive
definite. We define a real scalar product on the phase space $\phasesp$ and a
corresponding transposition using the bi-form $\gamma$
\begin{align}
  (\phi_1,\phi_2) = \phi_1^{\mathrm T}\circ\phi_2 =
  \phi_1\circ\gamma\circ\phi_2\qquad
  &\text{for}\quad \phi_1,\phi_2\in\phasesp\commae\nonumber\\
  A^{\mathrm T} = \gamma^{-1}\circ A\circ\gamma\qquad&\text{for}\quad
  A\in\phasesp^1_1\period\label{transp}
\end{align}

 From Eqs.\ (\ref{fiJJismdelta}), (\ref{fiJdFcompatibility}) follow
\begin{gather}
  M\circ \symplstr = - \symplstr\circ M \comma N\circ\symplstr = -
  \symplstr\circ N \commae\\
  -\delta = M\circ M - N\circ N + i (M\circ N + N\circ M)\spce
\end{gather}
i.e.
\begin{gather}
  N\circ N = \delta + M\circ M\commae \label{N2isdeltaplusM2}\\
  M\circ N = - N\circ M\period
\end{gather}

Now it is straightforward to show that $N$ is symmetric and $M$ is
antisymmetric with respect of the transposition (\ref{transp}). If we define
the absolute value $\abs{M}$ and signum $\sigma$ of the operator M
as\note{polardecomp}
\begin{align}
  \abs{M} &= (M^{\mathrm T}\circ M)^{\frac12} = (- M\circ
  M)^{\frac12}\commae\\
  \sigma &= \sign M = M\circ \abs{M}^{-1} = \abs{M}^{-1}\circ M\commae
\end{align}
we find
\begin{gather}
  \sigma^{\mathrm T} = \sigma^{-1}\comma \sigma^*=\sigma\comma
  \abs{M}^{\mathrm T} = \abs{M}\comma \abs{M}^* = \abs{M}\commae\\
  \sigma\circ\symplstr = - \symplstr\circ\sigma\comma
  \abs{M}\circ\symplstr = \symplstr\circ
  \abs{M}\commae\label{sigmadFcompatibility}\\
  \sigma\circ\sigma = -\delta\comma
  \sigma\circ\symplstr \quad\text{is positive definite}\commae
  \label{sigmadFpositivity}\\
  \bigl[\abs{M},N\bigr] = 0\comma
  \sigma\circ N = - N\circ\sigma\comma
  \sigma\circ\abs{M} = \abs{M}\circ\sigma\period
\end{gather}

Therefore $\abs{M}$ and $N$ have common eigenvectors and can be written as
functions of a single operator. We can find this operator solving Eq.\
(\ref{N2isdeltaplusM2}).
\begin{equation}
  N = \tanh \Chi\comma \abs{M} = \bigl[\cosh\Chi\bigr]^{-1}\period
\end{equation}
The operator $\Chi$ satisfies
\begin{gather}
  \bigl[N,\Chi\bigr] = 0 \comma \bigl[\abs{M},\Chi\bigr] = 0\commae\\
  \Chi^{\mathrm T} = \Chi\comma \Chi^* = \Chi\commae\\
  \Chi\circ\symplstr = -\symplstr\circ\Chi\comma
  \Chi\circ\sigma = - \sigma\circ\Chi \label{ChisigmadFcomrel}\period
\end{gather}
The Green function $\GFH$ can be written using the operators $\Chi$ and
$\sigma$ as
\begin{equation}
  \GFH\circ\symplstr = \sigma\circ\bigl[\cosh\Chi\bigr]^{-1} + i \tanh
  \Chi\period
\end{equation}
We will see below that the operator $\Chi$ is closely connected with
Bogoljubov transformation.

Finally we can express initial and final complex structures
\begin{equation} \label{JfiusingChisigma}
  J_i = \exp(-\Chi) \circ\sigma = \sigma\circ\exp(\Chi)\comma
  J_f = \exp(\Chi) \circ\sigma = \sigma\circ\exp(-\Chi)\commae
\end{equation}
and using this we see that $\Chi$ is $i$-antilinear, $f$-antilinear and
symmetric:
\begin{equation}
  \Chi\circ J_i = - J_i\circ\Chi\comma \Chi\circ J_f = - J_f\circ\Chi\comma
  \Chi^{\langle\top\rangle}= \Chi\period
\end{equation}

We have found that the Hadamard Green function $\GFH$ defined by Eq.\
(\ref{fiGFHdef}) satisfies
\begin{gather}
  F\bullet\GFH = 0 \comma \GFH{}^\top = \GFH\commae\label{GFHcondition1}\\
  \GFH\circ\symplstr\circ\GFH = \GFC\commae\label{GFHcondition2}\\
  \gamma = \bigl[\re\GFH\bigr]^{-1} \quad\text{is positive
  definite}\period\label{GFHcondition3}
\end{gather}
This is true also in an opposite direction. Any Green function which
satisfies these conditions is possible to write in the form (\ref{fiGFHdef}),
where the initial and final complex structures are given by
(\ref{JfiusingChisigma}).

We can also find boundary conditions for in-out Feynman Green function
similarly to Eq.\ (\ref{GFFboundcond}). Using Eqs.\ (\ref{oirelGFFtoMPH}),
(\ref{GFPMPiPfrel}) and (\ref{PMpartsboundcond}) we get
\begin{gather}
  -i\, (\chi[\domain]\delta) \bullet F \bullet \GFF = (\chi[\domain]\,\delta)
  \commae\nonumber\\
  (\Theta_f[\Sigma_f]^* - i \distdFr[\Sigma_f]) \bullet \GFF \bullet
  (\chi[\domain]\delta)   = 0 \commae\nonumber\\
  (\chi[\domain]\delta) \bullet\GFF\bullet (\Theta_f[\Sigma_f]^* - i
  \distdFl[\Sigma_f]) = 0 \commae \label{fiGFFboundcond}\\
  (\Theta_i[\Sigma_i] + i \distdFr[\Sigma_i]) \bullet \GFF \bullet
  (\chi[\domain]\delta) = 0 \commae\nonumber\\
  (\chi[\domain]\delta) \bullet\GFF\bullet (\Theta_i[\Sigma_i] + i
  \distdFl[\Sigma_i]) = 0 \spce\nonumber
\end{gather}
for any domain $\domain = \langle\Sigma_f|\Sigma_i\rangle$.

The relations among projectors $P^\pm, P^\pm_i, P^\pm_f$ and the operator $J$
can be translated to composition laws among Green function. Beside Eqs.\
(\ref{fiPPisP}), (\ref{fiJJismdelta}) we have, for example (using Eq.\
(\ref{relPifP}))
\begin{equation}
  \GFP_f\circ\symplstr\circ\GFP = \pm i\, \GFP\comma
  \GFP\circ\symplstr\circ\GFP_i = \mp i \GFP\period
\end{equation}


\subsection*{Bogoljubov operators}

We have studied the relation of two particle interpretations from the point
of view of Green functions. Now we will compare initial and final creation
and annihilation operators.

Because the structures of initial and final particle interpretations are
generated by creation and annihilation operators which satisfy the same
commutation relations, we can expect that they are related by a unitary
transformation. In other words, for any experiment formulated using the
initial notion of particles, we can construct an experiment formulated in the
same way using the final notion of particles. These experiments will be
generally different, but their description should be related by a unitary
transformation.

However this ``translation'' of the initial experiment to the final one is
not unique. We have to specify which initial and final states correspond each
other. We have to specify, for example, how we change modes which we use for
labeling of one-particle states.

More precisely, we write a relation of initial and final particle states in
following way
\begin{align}
  |f:vac\rangle &= \hat{S}^\dagger\, |i:vac\rangle\commae
  \label{Sfvacisivac}\\
  \hat{\mathrm a}_f[s\circ\phi]^\dagger |f:vac\rangle &= \hat{S}^\dagger \,
  \hat{\mathrm a}_i[\phi]^\dagger |i:vac\rangle \commae
  \label{Sf1partisi1part}
\end{align}
where $\hat{S}$ is a unitary operator on the quantum space $\quantsp$ called
the \emph{S-matrix},
\begin{equation}
  \hat{S}^\dagger = \hat{S}^{-1}
\end{equation}
and $s$ is a \emph{transition} operator on the phase space $\phasesp$ which
``translates'' initial modes to final modes as we discussed above. This means
that an initial one-particle state labeled by a mode $\phi$ is related by a
unitary transformation given by the S-matrix with a final one-particle state
labeled by the mode $s\circ\phi$. Of course, the S-matrix depends on a choice
of the operator $s$.

The relation (\ref{Sfvacisivac}), (\ref{Sf1partisi1part}) are equivalent to
\begin{equation}\label{fancrisSinacrS}
  \hat{\mathrm a}_f[s\circ\phi] = \hat{S}^\dagger \hat{\mathrm a}_i[\phi]
  \hat{S}\comma
  \hat{\mathrm a}_f[s\circ\phi]^\dagger = \hat{S}^\dagger \hat{\mathrm
  a}_i[\phi]^\dagger \hat{S}\period
\end{equation}
Using the commutation relation (\ref{ancrcomrel}) and unitarity of the
S-matrix we get a condition on the operator $s$,
\begin{equation}\label{fiscprrel}
  \langle s\circ\phi_1, s\circ\phi_2\rangle_f =
  \langle\phi_1,\phi_2\rangle_i\period
\end{equation}
The operator $s$ changes the initial scalar product on the phase space
$\phasesp$ to the final one. This is a natural condition which expresses a
meaning of the transition operator $s$ --- this operator translates the
initial labeling of one-particle states to the final one, and therefore it
has to map all initial structures on the phase space to the final ones.
Consequences of the last equation are
\begin{equation}\label{sproperties}
  s\circ\gamma_f\circ s = \gamma_i\comma
  s\circ\symplstr\circ s = \symplstr\comma
  J_f \circ s = s \circ J_i\period
\end{equation}
Using these relations and Eq.\ (\ref{fancrisSinacrS}) we can get another
relation between the S-matrix and the operator $s$,
\begin{equation}
  s\circ\qPhi = \hat{S} \qPhi \hat{S}^\dagger\period
\end{equation}

The transition operator $s$ is closely related to Bogoljubov coefficients
between initial and final bases of modes. It can be seen from a decomposition
of $s$ to $i$-linear and $i$-antilinear parts (see Eqs.\ (\ref{plinearity}),
(\ref{pantilinerity}))
\begin{gather}
  s = \alpha + \beta\commae\label{sisaplusb}\\
  \alpha\circ J_i = J_i \circ\alpha \comma \beta \circ J_i = - J_i \circ
  \beta\period
\end{gather}
Explicitly
\begin{equation}\label{alphabetadef}
\begin{split}
  \alpha &= \frac12 ( s - J_i\circ s\circ J_i) = P^+_i\circ s \circ P^+_i +
  P^-_i\circ s \circ P^-_i\commae\\
  \beta &= \frac12 ( s + J_i\circ s\circ J_i) = P^+_i\circ s \circ P^-_i +
  P^-_i\circ s \circ P^+_i\period
\end{split}
\end{equation}
This allows us to write relations of initial and final positive and negative
frequency projectors,
\begin{equation}\label{Pfirelusingab}
\begin{split}
  P^+_f\circ s &= P^+_i\circ\alpha + P^-_i\circ\beta\commae\\
  P^-_f\circ s &= P^-_i\circ\alpha + P^+_i\circ\beta\spce
\end{split}
\end{equation}
and relations of initial and final creation and annihilation operators,
\begin{equation}\label{finacrrelusingab}
\begin{split}
  \hat{\mathrm a}_f[s\circ\phi] &= \hat{\mathrm a}_i[\alpha\circ\phi] -
  \hat{\mathrm a}_i[\beta\circ\phi]^\dagger\commae\\
  \hat{\mathrm a}_f[s\circ\phi]^\dagger &= \hat{\mathrm
  a}_i[\alpha\circ\phi]^\dagger - \hat{\mathrm a}_i[\beta\circ\phi]\period
\end{split}
\end{equation}
This expresses final creation resp. annihilation operators as a mixture of
both initial creation and annihilation operators. It shows explicitly that
the initial and final notion of particles are really different for
$\beta\not=0$. The relations (\ref{finacrrelusingab}) are a base-independent
definition of the Bogoljubov transformation, and we will call the operators
$\alpha, \beta$ \emph{Bogoljubov operators}.

Now we will investigate properties of Bogoljubov operators. It is
straightforward to show (see Eqs.\ (\ref{phermconj}), (\ref{ptranspos}) and
(\ref{prealtranspos}) for definition of used operations)
\begin{gather}
  s^{-1} = - J_i \circ s^{{\mathrm T}_i}\circ J_i =
  \alpha^{\langle\dagger\rangle} - \beta^{\langle\top\rangle}\commae\\
  s\circ s^{{\mathrm T}_i} = - J_f\circ J_i\period
\end{gather}
Substituting this and Eq.\ (\ref{sisaplusb}) to $\delta = s\circ s^{-1} =
s^{-1}\circ s$ and taking $i$-linear and $i$-antilinear parts we get
identities
\begin{gather}
  \alpha\circ\alpha^{\langle\dagger\rangle} -
  \beta\circ\beta^{\langle\top\rangle} = \delta\comma
  \alpha^{\langle\dagger\rangle}\circ \alpha -
  \beta^{\langle\top\rangle}\circ\beta = \delta\commae\\
  \beta\circ\alpha^{\langle\dagger\rangle} = \alpha\circ
  \beta^{\langle\top\rangle}\comma
  \alpha^{\langle\dagger\rangle}\circ\beta = \beta^{\langle\top\rangle}\circ
  \alpha\period
\end{gather}

The transition operator $s$ is not fixed by Eq.\ (\ref{fiscprrel}) uniquely,
we have a freedom in selecting this operator. It corresponds to a freedom in
labeling of our one-particle states. We can change the labeling of initial
one-particle states by $i$-unitary transformation of the phase space without
changing the initial notion of particles and similarly with the final
particles. If we have two different transition operators $s_a$ and $s_b$ for
translation of initial modes to final modes, they both have to satisfy Eq.\
(\ref{fiscprrel}) and we easily see that they have to be related by
\begin{equation}\label{ssrel}
  s_a = s_b\circ u_i = u_f\circ s_b  \commae
\end{equation}
where $u_i$ resp. $u_f$ is an $i$-linear and $i$-unitarian resp. $f$-linear
and $f$-unitarian operator on the phase space $\phasesp$ (see Eqs.\
(\ref{plinearity}), (\ref{phermconj})).

All equations can be simplified if we choose a special transition operator
$s$. The operator
\begin{equation}
  s_o = \exp \Chi = - \sigma\circ J_i = - J_f \circ \sigma = \bigl[-J_f\circ
  J_i\bigr]^{\frac12}
\end{equation}
satisfies conditions (\ref{sproperties}) and therefore also the condition
(\ref{fiscprrel}), and so we can use it as a special transition operator for
translation of initial to final modes. We will call this choice
\emph{canonical}. We can define canonical Bogoljubov operators $\alpha_o,
\beta_o$ by Eq.\ (\ref{alphabetadef}) and using Eq.\ (\ref{sproperties}), we
get
\begin{gather}
  s_o = \exp(\Chi) = \alpha_o + \beta_o\comma
  s_o^{-1} = \exp(-\Chi) = \alpha_o - \beta_o\commae\\
  \alpha_o = \cosh \Chi\comma \beta_o = \sinh \Chi\commae\\
  \alpha_o\circ J_{i,f} = J_{i,f}\circ \alpha_o\comma
  \beta_o\circ J_{i,f} = - J_{i,f}\circ\beta_o\commae\\
  s_o^{{\mathrm T}_{i,f}} = s_o\comma
  \alpha_o^{\langle\dagger\rangle} = \alpha_o \comma
  \beta_o^{\langle\top\rangle} = \beta_o\period
\end{gather}
We also see that the operators $\alpha_o$ and $-\beta_o$ play a role of
inverse canonical Bogoljubov operators for transformation from final to
initial particle states.

This is possible to generalize for any transition operator $s$. Using Eq.\
(\ref{ssrel}) we get
\begin{gather}
\begin{split}
  s = \exp(\Chi) \circ u_i &= u_f\circ\exp(\Chi) = \alpha + \beta\commae\\
  s^{-1} = \exp(-\Chi) \circ u_i &= u_f\circ\exp(-\Chi) =
  \alpha^{\langle\dagger\rangle} - \beta^{\langle\top\rangle}\commae
\end{split}\\
  \alpha\circ J_{i,f} = J_{i,f}\circ \alpha\comma \beta\circ J_{i,f} = -
  J_{i,f}\circ\beta  \period
\end{gather}
We see that the inverse Bogoljubov operators are
$\alpha^{\langle\dagger\rangle}$ and $-\beta^{\langle\top\rangle}$. Finally,
let us note that
\begin{equation}
  \beta\circ\alpha^{-1} = \alpha^{-1}\circ\beta = \tanh\Chi = N\period
\end{equation}

Now we will show a connection with standard Bogoljubov coefficients. Let's
assume that a $i$-orthonormal resp $f$-orthonormal bases ${\mathbf u} =
\{u_{\mathrm k}; {\mathrm k}\in {\mathcal I}\}$ resp. ${\mathbf v} =
\{v_{\mathrm k}; {\mathrm k}\in {\mathcal I}\}$ for  description of initial
resp. final modes are chosen. It defines the transition operator $s$ by the
conditions
\begin{equation}\label{vkissuk}
  v_{\mathrm k} = s\circ u_{\mathrm k}\comma J_f\circ v_{\mathrm k} = s \circ
  J_i \circ u_{\mathrm k}
\end{equation}
for all ${\mathrm k}\in {\mathcal I}$. The Bogoljubov coefficients
$\alpha_{\mathrm {kl}}, \beta_{\mathrm {kl}}$ are defined by the equations
\cite{DeWitt:1975}
\begin{equation}
  v_{\mathrm k}\frqp{+}{f} = \sum_{{\mathrm k}\in {\mathcal I}}
  \bigl(u_{\mathrm l}\frqp{+}{i} \alpha_{\mathrm {lk}}  + u_{\mathrm
  l}\frqp{-}{i} \beta_{\mathrm {lk}} \bigr)\period
\end{equation}
Using Eq.\ (\ref{Pfirelusingab}) we see
\begin{equation}
  \alpha_{\mathrm {kl}} = \langle u_{\mathrm k}, \alpha\circ u_{\mathrm
  l}\rangle_i\comma
  \beta_{\mathrm {kl}} = \langle u_{\mathrm k}, \beta\circ u_{\mathrm
  l}\rangle_i\commae
\end{equation}
i.e. Bogoljubov coefficients are matrix elements of the Bogoljubov operators
in a chosen base.

We can use eigenvectors of the operator $\Chi$ to define special bases of
initial and final modes --- so called canonical bases \cite{Hajicek:1977}.
Because the operator $\Chi$ is $i$-symmetric it has a complete set of
eigenvectors. From $i$-antilinearity follows that for each eigenvector $u$
the vector $J_i\circ u$ is also an eigenvector with the opposite sign of the
eigenvalue. Therefore we can choose an ${\mathbb R}$-base $\{u_{\mathrm k},
J_i\circ u_{\mathrm k}; {\mathrm k}\in {\mathcal I}\}$ such that
\begin{equation}
\begin{split}
  \Chi\circ u_{\mathrm k} &= \chi_{\mathrm k}\, u_{\mathrm k}\quad,\qquad
  \chi_{\mathrm k} \geq 0 \commae\\
  \Chi\circ J_i\circ u_{\mathrm k} &= - \chi_{\mathrm k}\, J_i\circ
  u_{\mathrm k}\period
\end{split}
\end{equation}
The base can be chosen orthonormal with respect to the real scalar product
defined by the bi-form $\gamma_i$
\begin{equation}
  u_{\mathrm k}\circ\gamma_i\circ u_{\mathrm l} = \delta_{\mathrm {kl}}\comma
  u_{\mathrm k}\circ\gamma_i\circ J_i\circ u_{\mathrm l} = 0\period
\end{equation}
If the operator $\Chi$ is nondegenerate with different eigenvalues, the base
is fixed uniquely. The subset $\{u_{\mathrm k}; {\mathrm k}\in {\mathcal
I}\}$ of this ${\mathbb R}$-base forms the $i$-orthonormal ${\mathbb
C}$-base. It will be used for labeling of initial particles. The final
$f$-orthonormal ${\mathbb C}$-base $\{v_{\mathrm k}; {\mathrm k}\in {\mathcal
I}\}$ will be generated by the canonical transition operator $s_o$,
\begin{equation}
  v_{\mathrm k} = s_o \circ u_{\mathrm k} = \exp(\chi_{\mathrm
  k})\,u_{\mathrm k}\period
\end{equation}
The Bogoljubov transformation between these two bases is
\begin{equation}
  v_{\mathrm k}\frqp{+}{f} = u_{\mathrm k}\frqp{+}{i} \cosh(\chi_{\mathrm k})
  + u_{\mathrm k}\frqp{-}{i} \sinh(\chi_{\mathrm k})
\end{equation}
and the final and initial creation and annihilation operators are related by
\begin{equation}
\begin{split}
  \hat{\mathrm a}_f[v_{\mathrm k}]
  &= \hat{\mathrm a}_i[u_{\mathrm k}] \cosh(\chi_{\mathrm k}) - \hat{\mathrm
  a}_i[u_{\mathrm k}]^\dagger \sinh(\chi_{\mathrm k})\commae\\
  \hat{\mathrm a}_f[v_{\mathrm k}]^\dagger
  &= \hat{\mathrm a}_i[u_{\mathrm k}]^\dagger \cosh(\chi_{\mathrm k}) -
  \hat{\mathrm a}_i[u_{\mathrm k}] \sinh(\chi_{\mathrm k})\period
\end{split}
\end{equation}


\subsection*{Transition amplitudes}

Now we are prepared to calculate in-out transition amplitudes, i.e.
amplitudes between initial and final particle states. It is possible to show
(see \cite{Krtous:pigflong,DeWitt:1975}) that amplitudes between many
particle states can be reduced to one particle transition amplitudes.
Therefore we will calculate only these simple amplitudes. First, using Eq.\
(\ref{finacrrelusingab}) and its inversion, we get identities
\begin{gather}
  \hat{\mathrm a}_i[\phi]^\dagger = \hat{\mathrm
  a}_f[s\circ\alpha^{-1}\circ\phi]^\dagger +
  \hat{\mathrm a}_i[\beta\circ\alpha^{-1}\circ\phi]\commae\nonumber\\
  \hat{\mathrm a}_f[\phi] = \hat{\mathrm
  a}_i[s^{-1}\circ\alpha^{-1\,{\langle\dagger\rangle}}\circ\phi] -
  \hat{\mathrm
  a}_f[\beta^{\langle\top\rangle}\circ\alpha^{-1\,{\langle\dagger\rangle}}%
\circ\phi]^\dagger\period \label{fiancrrelrev}
\end{gather}

Now it is easy to check that vacuum -- one-particle transition amplitudes
vanish
\begin{equation}
\begin{split}
  \langle f:vac|\hat{\mathrm a}_i[\phi]^\dagger |i:vac\rangle &=
  \langle f:vac|\bigl(\hat{\mathrm a}_f[s\circ\alpha^{-1}\circ\phi]^\dagger
  + \hat{\mathrm a}_i[\beta\circ\alpha^{-1}]\bigr)|i:vac\rangle = 0\commae\\
  \langle f:vac|\hat{\mathrm a}_f[\phi]|i:vac\rangle &= 0\period
\end{split}
\end{equation}
The one-particle to one-particle transition amplitude is
\begin{equation}
\begin{split}
  &\frac{\langle f:vac|\hat{\mathrm a}_f[s\circ\phi_1]\hat{\mathrm
  a}_i[\phi_2]^\dagger |i:vac\rangle}{\langle f:vac|i:vac\rangle} =\\
  &\quad= \frac{\langle f:vac|\hat{\mathrm
  a}_f[s\circ\phi_1]\bigl(\hat{\mathrm
  a}_f[s\circ\alpha^{-1}\circ\phi_2]^\dagger +
  \hat{\mathrm a}_i[\beta\circ\alpha^{-1}\circ\phi_2]\bigr)
  |i:vac\rangle}{\langle f:vac|i:vac\rangle} =\\
  &\quad= \langle s\circ\phi_1,s\circ\alpha^{-1}\circ\phi_2\rangle_f =
  \langle\phi_1,\alpha^{-1}\circ\phi_2\rangle_i = \phi_1 \circ
  \traI\circ\phi_2  \commae
\end{split}
\end{equation}
where we have used the commutation relation (\ref{ancrcomrel}). Vacuum to
two-particle transition amplitudes are
\begin{align}
\begin{split}
  &\frac{\langle f:vac|\hat{\mathrm a}_i[\phi_1]^\dagger \hat{\mathrm
  a}_i[\phi_2]^\dagger|i:vac\rangle}{\langle f:vac|i:vac\rangle} =\\
  &\quad= \frac{\langle f:vac|\bigl(\hat{\mathrm
  a}_f[s\circ\alpha^{-1}\circ\phi_1]^\dagger +
  \hat{\mathrm a}_i[\beta\circ\alpha^{-1}\circ\phi_1]\bigr) \hat{\mathrm
  a}_i[\phi_2]^\dagger |i:vac\rangle}{\langle f:vac|i:vac\rangle} =\\
  &\quad= \langle\beta\circ\alpha^{-1}\circ\phi_1,\phi_2\rangle_i =
  \phi_1\circ\traL\circ\phi_2\commae
\end{split}\\
  &\frac{\langle f:vac|\hat{\mathrm a}_f[s\circ\phi_1] \hat{\mathrm
  a}_f[s\circ\phi_2] |i:vac\rangle}{\langle f:vac|i:vac\rangle} =
  - \langle\phi_1, \alpha^{-1}\circ\beta\circ\phi_2\rangle_i = \phi_1\circ
  \traV \circ \phi_2\period
\end{align}
We have introduced amplitudes $\traI, \traV, \traL$ similarly to De Witt's
notation \cite{DeWitt:1975}
\begin{align}
  \traI &= -i\,P^-_i\circ\symplstr\circ\alpha^{-1}\circ P^+_i
  \commae\nonumber\\
  \traL &= i\,P^+_i\circ\symplstr\circ \tanh(\Chi)\circ P^+_i\comma
  \traL^\top = \traL\commae \label{VLIdef}\\
  \traV &= i\,P^-_i\circ\symplstr\circ \tanh(\Chi)\circ P^-_i\comma
  \traV^\top = \traV\period\nonumber
\end{align}
The bi-form $\traV$, and $\traL$ are independent of a choice of the operator
$s$, and $\traI$ is $s$-dependent.

These expressions are connected more closely with initial modes --- we have
used a transition operator $s$ to generate final modes. It is possible to
also obtain translation amplitudes without this asymmetry. Simply choosing
the canonical operator $s_o$ and doing some algebra, we can get
\begin{equation}
\begin{split}
  &\frac{\langle f:vac|\hat{\mathrm a}_f[\phi_1]\hat{\mathrm
  a}_i[\phi_2]^\dagger |i:vac\rangle}{\langle f:vac|i:vac\rangle} =
  \langle(\delta - \tanh\Chi)\circ\phi_1,\phi_2\rangle_i =
  \langle\phi_1,(\delta+\tanh\Chi)\circ\phi_2\rangle_f =\\
  &\qquad = - \phi_1\circ\symplstr\circ\GFP\circ\symplstr\circ\phi_2 \commae
\end{split}
\end{equation}
\begin{align}
\begin{split}
  &\frac{\langle f:vac|\hat{\mathrm a}_i[\phi_1]^\dagger \hat{\mathrm
  a}_i[\phi_2]^\dagger |i:vac\rangle}{\langle f:vac|i:vac\rangle} =
  \langle\tanh(\Chi)\circ\phi_1,\phi_2\rangle_i =\\
  &\qquad = \frac12\, \phi_1\circ P^+_i
  \circ\symplstr\circ\GFH\circ\symplstr\circ P^+_i\circ\phi_2 \commae
\end{split}\\
\begin{split}
  &\frac{\langle f:vac|\hat{\mathrm a}_f[\phi_1]\hat{\mathrm
  a}_f[\phi_2]|i:vac\rangle}{\langle f:vac|i:vac\rangle} =
  - \langle\phi_1,\tanh(\Chi)\circ\phi_2\rangle_f = \\
  &\qquad = \frac12 \phi_1\circ P^-_f
  \circ\symplstr\circ\GFH\circ\symplstr\circ P^-_f\circ\phi_2 \period
\end{split}
\end{align}

We have thus calculated the in-out one-particle transition amplitudes, except
for calculating the normalization factor $\langle f:vac|i:vac\rangle$. In
\cite{Krtous:pigflong} (see also \cite{DeWitt:1975}) it is derived that
\begin{equation}\label{vacvacampl}
  \langle f:vac|i:vac\rangle = \bigl(\detix_{\phasesp}\,\cosh
  \Chi\bigr)^{-\frac14} =
  \bigl(\detix_{\phasesp_{i,f}}\,\cosh \Chi\bigr)^{-\frac12} =
  \biggl(\prod_{{\mathrm k}\in {\mathcal I}} \cosh \chi_{\mathrm
  k}\biggr)^{-\frac12}\period
\end{equation}

Other interesting physical quantities are the mean number of final particles
in the initial vacuum. Using the identities (\ref{finacrrelusingab}) we get
\begin{equation}
  \langle i:vac|\hat{\mathrm n}_f[s\circ\phi]|i:vac\rangle =
  \frac{\langle\beta\circ\phi,\beta\circ\phi\rangle_i}{\langle\phi,\phi%
\rangle_i} =
  \frac{\langle\sinh(\Chi)\circ\phi,\sinh(\Chi)\circ\phi\rangle_i}{\langle%
\phi,\phi\rangle_i}
\end{equation}
or
\begin{equation}
  \langle i:vac|\hat{\mathrm n}_f[\phi]|i:vac\rangle =
   \frac{\langle\sinh(\Chi)\circ\phi,\sinh(\Chi)\circ\phi\rangle_f}{\langle%
\phi,\phi\rangle_f}\spce
\end{equation}
and similarly
\begin{equation}
  \langle f:vac|\hat{\mathrm n}_i[\phi]|f:vac\rangle =
   \frac{\langle\sinh(\Chi)\circ\phi,\sinh(\Chi)\circ\phi\rangle_i}{\langle%
\phi,\phi\rangle_i}\period
\end{equation}
The mean total number of particles is
\begin{equation}
  \langle i:vac|\hat{\mathrm N}_f|i:vac\rangle =
  \sum_{{\mathrm k}\in {\mathcal I}} \langle i:vac|\hat{\mathrm
  n}_f[v_{\mathrm k}]|i:vac\rangle =
  \sum_{{\mathrm k}\in {\mathcal I}} \langle v_{\mathrm k},
  (\sinh\Chi)^2\circ v_{\mathrm k}\rangle_f
\end{equation}
for some $f$-orthonormal ${\mathbb C}$-base, i.e.
\begin{equation}
  \langle i:vac|\hat{\mathrm N}_f|i:vac\rangle = \Tr_{\phasesp_f}
  (\sinh\Chi)^2 =
  \sum_{{\mathrm k}\in {\mathcal I}} (\sinh\chi_{\mathrm k})^2\spce
\end{equation}
and similarly
\begin{equation}
  \langle f:vac|\hat{\mathrm N}_i|f:vac\rangle = \Tr_{\phasesp_i}
  (\sinh\Chi)^2 =
  \sum_{{\mathrm k}\in {\mathcal I}} (\sinh\chi_{\mathrm k})^2\period
\end{equation}
The regularity of this quantity guarantees the unitary equivalence of the
initial and final particle representations of the quantum algebra and
regularity of the vacuum -- vacuum amplitude (\ref{vacvacampl}) (see
\cite{Wald:1979a,Wald:book1994}).


 \newpage
 \section*{Acknowledgments}

I would like to thank Professor D. N. Page for his support during this work
and careful reading of the manuscript and Tom\'a\v s Kopf for useful
discussions. The financial support has been provided in part by the Natural
Sciences and Engineering Council of Canada.


\begin{notes}
\noteitem{bundlesnotation}{
  We will denote $\fctsct\,M$ the space of the functions on manifold $M$,
  $\dnstsct^{\alpha}\,M$ the space of densities on $M$ of a weight $\alpha$.
  If $V$ is a vector space and $V^*$ its dual, $V^k_l$ will denote the tensor
  space
  \[\underbrace{V\otimes\dots\otimes V}_{k\text{
  times}}\otimes\underbrace{V^*\otimes\dots\otimes V^*}_{l\text{
  times}}\period
  \]}
\noteitem{positasindex}{
  $\phi$ means an abstract vector from $\hist$ with coordinates
  $\Phi^{x}(\phi) = \phi^{x} = \phi(x)$. The coordinates of a covector
  $\omega_{x} = \omega(x)$ are densities on the manifold $M$.}
\noteitem{indexes}{
  We are using MTW \cite{MTW} signs conventions and geometric units $c=1,
  \hbar=1$. Greek indexes ($\alpha$, $\beta$, $\gamma, \dots$) are used for
  spacetime tensors, latin letters from the beginning of alphabet ($a$, $b$,
  $c, \dots$) for space tensors and from the end of alphabet ($\dots, x$,
  $y$, $z$) for spacetime points. In rare occasions when we will need indexes
  for vectors from $\phasesp$ we will use also letters $..., x$, $y$, $z$.
  Bold indexes represent abstract vector indexes in the sense of Penrose
  \cite{PenroseRindler:book}. Plain indexes are coordinate indexes. }
\noteitem{distributions}{
  The delta distribution $\delta^x_{x'} = \delta(x|x')$ is a density in one
  point and function in other one, $(\gdnst \delta)_{xx'}$ is a density in
  both points and $(\chi[\domain]\,\delta)^x_{x'}$ is a projection on the
  domain $\domain$ (i.e. $(\chi[\domain]\,\delta) \bullet \psi =
  \chi[\domain]\,\psi$).
  The distributions $\distdl_{\absidx{\alpha}}, \distdr_{\absidx{\alpha}}$
  are distributions representing gradients of delta function defined by
  \begin{gather*}
    \varphi^{\absidx{\alpha}} \bullet \distdr_{\absidx{\alpha}} \bullet \psi
    =
    \psi \bullet \distdl_{\absidx{\alpha}} \bullet \varphi^{\absidx{\alpha}}
    =
    \int{\varphi^{\absidx{\alpha}} (\grad_{\absidx{\alpha}} \psi)}\\
    \varphi^{\absidx{\alpha}} \bullet \distdl_{\absidx{\alpha}} \bullet \psi
    =
    \psi \bullet \distdr_{\absidx{\alpha}} \bullet \varphi^{\absidx{\alpha}}
    =
    \int{(\grad_{\absidx{\alpha}} \varphi^{\absidx{\alpha}}) \psi}
  \end{gather*}
  for any test function $\psi\in\fctsct\,M$ and test vector density
  $\varphi^{\absidx{\alpha}}$. Similiarly $\Boxr$ is the d'Alembert operator
  acting on delta function.}
\noteitem{operatorsconv}{
  We are using following convention for operators ($A\in\phasesp^1_1,
  \phi\in\phasesp, \omega\in\phasesp^0_1$)
  \begin{gather*}
    \omega\circ A\circ\phi =
    \omega_{\absidx{x}}\,
    A^{\absidx{x}}_{\absidx{y}}\,\phi^{\absidx{y}}\commae\\
    \phi\circ A\circ\omega =
    \phi^{\absidx{y}}\, A^{\absidx{x}}_{\absidx{y}}\,
    \omega_{\absidx{x}}\period
  \end{gather*}
  It means that in the first case the operator $A$ acts to the right, in the
  later case to the left. The order is determined by the fact that $\phi$ is
  vector and $\omega$ is covector. This convention simulate contraction of
  vector indexes and it is necessary to be careful in some cases. For example
  if $A,B$ are operators on $\phasesp$ we can write
  \begin{equation*}
    (A\circ\symplstr \circ \GFC \circ B)^{\absidx{x}}_{\absidx{y}} =
    A^{\absidx{u}}_{\absidx{y}} \symplstr_{\absidx{uz}} \GFC^{\absidx{zv}}
    B^{\absidx{x}}_{\absidx{v}} =
    - A^{\absidx{u}}_{\absidx{y}}
    {\delta^{(\phasesp)}}^{\absidx{v}}_{\absidx{u}}
    B^{\absidx{x}}_{\absidx{v}} =
    - (B\circ A)^{\absidx{x}}_{\absidx{y}} \not= - (A\circ
    B)^{\absidx{x}}_{\absidx{y}}\period
  \end{equation*}}
\noteitem{gammavariat}{
  Here the ``gradient'' $\variatb$ acts on the phase space $\phasesp$.}
\noteitem{onlybosons}{
  Because for the description of a scalar field we need bosonic particles, we
  are assuming only the bosonic version of Fock space. }
\noteitem{phasespdelta}{
  More precisely we should use here the symbol $\delta^{(\phasesp)}$ for the
  identity operator on $\phasesp$ as we did before. We will omit a
  superscript if it is clear from context that we are speaking about
  operators on $\phasesp$.}
\noteitem{nonhomGFeq}{
  We cannot write simply
  \begin{equation*}
    -i F \bullet \GFF_p = \delta \tag{*}
  \end{equation*}
  (similarly for $\GFR, \GFA$ and $\GFS$) because for $\phi\in\phasesp$ we
  would get
  \begin{equation*}\label{contradiction}
    \phi = \phi\bullet\delta = \phi\bullet (-i) F\bullet \GFF_p = 0 \tag{**}
  \end{equation*}
  using the equation of motion (\ref{freeeqofmot}). F has zero eigenvectors
  (exactly the phase space $\phasesp$) and therefore it does not have an
  inverse tensor.

  We could use Eq.\ (*) if we understood $F_{xz} {\GFF_P}^{zy}$ as a
  distribution in the argument $x$ acting only an test functions with compact
  support. The step function $\chi[\domain]$ supply this conditions even for
  $\phi\in\phasesp$ which does not have a compact support, but it produces
  new important boundary terms and therefore the improved equation
  (\ref{eqforGFF}) does not lead to the contradiction (**)
  \begin{equation*}
  \begin{split}
    &\phi\bullet(\chi[\domain]\,\delta) =
    -i\,\phi\bullet(\chi[\domain]\,\delta)\bullet F\bullet\GFF_p =\\
    &\quad= -i\,\phi \bullet F \bullet (\chi[\domain]\,\delta) \bullet \GFF_p
    - i\,\phi \bullet \symplstr[\partial \domain] \bullet\GFF_p =\\
    &\quad= -i\,\phi \bullet
    \symplstr[\Sigma_f]\bullet\GFP\bullet(\chi[\domain]\,\delta)
    +
    i\,\phi\bullet\symplstr[\Sigma_i]\bullet\GFM\bullet(\chi[\domain]\,%
\delta) =\\
    &\quad=  \phi \bullet \symplstr[\Sigma] \bullet
    \GFC\bullet(\chi[\domain]\,\delta) = \phi\bullet(\chi[\domain]\,\delta)
  \end{split}
  \end{equation*}
  where we have used equations (\ref{eqforGFF}), (\ref{Fchicomm}),
  (\ref{relGFFtoMPH}), (\ref{relGFHCtoPM}) and (\ref{defGFCinphasesp}). }
\noteitem{polardecomp}{
  Polar decomposition is a decomposition of an operator in a Hilbert space
  into its absolute value and signum. We will use it for a real Hilbert
  space, i.e. a real vector space with a scalar product defined by a
  symmetric positive quadratic form $h$
  \begin{gather*}
    (a,b) = a^{\mathbf T} \cdot b = a \cdot h \cdot b \qquad \text{$a,b$
    vectors}\commae\\
    A^{\mathbf T} = h^{-1}\cdot A \cdot h \qquad \text{$A$ an
    operator}\period
  \end{gather*}
  There exists unique left resp. right decomposition of an operator $O$
  \[
    O = \abs{O}_l \cdot \sign_l O = \sign_r O\cdot \abs{O}_r
  \]
  to a positive definite symmetric operator $\abs{O}_l$ resp. $\abs{O}_r$ and
  an orthogonal operator $\sign_l O$ resp. $\sign_r O$
  \begin{gather*}
    \abs{O}_{l,r}^{\mathbf T} = \abs{O}_{l,r} \comma \abs{O}_{l,r}
    \quad\text{positive definite}\commae\\
    (\sign_{l,r} O)^{\mathbf T} = (\sign_{l,r} O)^{-1}\spce
  \end{gather*}
  and these operators are given by
  \begin{gather*}
    \abs{O}_l = \bigl( O\cdot O^{\mathbf T}\bigr)^{\frac12}\comma \abs{O}_r =
    \bigl( O^{\mathbf T}\cdot O\bigr)^{\frac12}\commae\\
    \sign_l O = \abs{O}_l^{-1}\cdot O\comma \sign_r O =
    O\cdot\abs{O}_r^{-1}\period
  \end{gather*}
  If $O$ commutes with $O^{\mathbf T}$ both decompositions coincide.}
\end{notes}



\end{document}